\documentclass[11pt,a4paper]{article}
\usepackage{amssymb}
\usepackage[dvips]{graphicx}
\usepackage{bm}
\usepackage{amsmath, textcomp}
\begin{document}

\title{Two-loop renormalization of the matter superfields and finiteness of ${\cal N}=1$ supersymmetric gauge theories regularized by higher derivatives}

\author{
A.E.Kazantsev$^1$, K.V.Stepanyantz$^2$\\
{\small{\em Moscow State University}}, {\small{\em  Faculty of Physics}},\\ {\small{\em $^1$Department of Quantum Theory and High Energy Physics}} \\{\small{\em $^2$Department  of Theoretical Physics}}\\
{\small{\em 119991, Moscow, Russia}}}

\maketitle

\begin{abstract}
The two-loop anomalous dimension of the chiral matter superfields is calculated for a general ${\cal N}=1$ supersymmetric gauge theory regularized by higher covariant derivatives. We obtain both the anomalous dimension defined in terms of the bare couplings, and the one defined in terms of the renormalized couplings for an arbitrary renormalization prescription. For the one-loop finite theories we find a simple relation between the higher derivative regulators under which the anomalous dimension defined in terms of the bare couplings vanishes in the considered approximation. In this case the one-loop finite theory is also two-loop finite in the HD+MSL scheme. Using the assumption that with the higher covariant derivative regularization the NSVZ equation is satisfied for RGFs defined in terms of the bare couplings, we construct the expression for the three-loop $\beta$-function. Again, the result is written both for the $\beta$-function defined in terms of the bare couplings and for the one defined in terms of the renormalized couplings for an arbitrary renormalization prescription.
\end{abstract}

\section{Introduction}
\hspace*{\parindent}

Quantum properties of supersymmetric theories have a lot of very interesting features. The maximally extended ${\cal N}=4$ supersymmetric Yang--Mills theory (SYM) is finite in all loops \cite{Grisaru:1982zh,Mandelstam:1982cb,Brink:1982pd,Howe:1983sr}, while ${\cal N}=2$ supersymmetric gauge theories are finite beyond the one-loop approximation \cite{Grisaru:1982zh,Howe:1983sr,Buchbinder:1997ib}.  This implies that ${\cal N}=2$ theories finite in the one-loop approximation are also finite in all orders \cite{Howe:1983wj}. The one-loop finiteness can be achieved by a special choice of a gauge group and a representation for the matter superfields. In ${\cal N}=1$ supersymmetric theories the divergent quantum corrections can exist in all orders. However, supersymmetry leads to some interesting relations between various renormalization constants. For example, due to the finiteness of the superpotential \cite{Grisaru:1979wc} the renormalizations of masses and Yukawa couplings are related to the renormalization of the chiral matter superfields. Similarly, the non-renormalization of the triple gauge-ghost vertices \cite{Stepanyantz:2016gtk} allows choosing a renormalization prescription for which

\begin{equation}\label{Z_Triple_Relation}
Z_\alpha^{-1/2} Z_c Z_V = 1,
\end{equation}

\noindent
where $Z_\alpha$, $Z_c$, and $Z_V$ are the renormalization constants for the gauge coupling constant, the Faddeev--Popov ghosts, and the quantum gauge superfield, respectively. However, the most interesting quantum feature of ${\cal N}=1$ supersymmetric gauge theories is the existence of a relation between the $\beta$-function and the anomalous dimension of the matter superfields called ``the exact NSVZ $\beta$-function'' \cite{Novikov:1983uc,Jones:1983ip,Novikov:1985rd,Shifman:1986zi}. Usually it is written as\footnote{In this section we do not specify the definition of the renormalization group functions. This will be done below.}

\begin{equation}\label{NSVZ_Equation_Old}
\beta(\alpha,\lambda) = -\frac{\alpha^2(3C_2 - T(R) + C(R)_i{}^j (\gamma_\phi)_j{}^i(\alpha,\lambda)/r)}{2\pi(1-C_{2}\alpha/(2\pi))},
\end{equation}

\noindent
where $r$ is the dimension of the gauge group $G$ with the structure constants $f^{ABC}$, and the group Casimirs are defined as

\begin{equation}
f^{ABC} f^{ABD} = \delta^{CD} C_2; \qquad \mbox{tr}(T^A T^B)= T(R) \delta^{AB};\qquad C(R)_i{}^j = (T^A T^A)_i{}^j.
\end{equation}

\noindent
In our notation $T^A$ are the generators of the representation to which the matter superfields belong. They should be distinguished from the generators of the fundamental representation denoted by $t^A$, which are assumed to be normalized by the condition $\mbox{tr}(t^A t^B) = \delta^{AB}/2$.

If ${\cal N}=2$ supersymmetric gauge theories are considered as a particular case of ${\cal N}=1$ theories, then the NSVZ relation leads to the finiteness beyond the one-loop approximation, provided the quantization is manifestly ${\cal N}=2$ supersymmetric \cite{Shifman:1999mv,Buchbinder:2014wra}. A natural way to provide this is to use harmonic superspace \cite{Galperin:1984av,Galperin:2001uw} and an invariant regularization \cite{Buchbinder:2015eva}. In this case we will also obtain the all-loop finiteness of ${\cal N}=4$ SYM theory as a consequence of the NSVZ equation.

It is important to recall that the NSVZ equation is valid only for certain renormalization prescriptions which are usually called ``the NSVZ schemes''. In particular, the most popular $\overline{\mbox{DR}}$ scheme (which is obtained in the case of using dimensional reduction \cite{Siegel:1979wq} supplemented by modified minimal subtraction \cite{Bardeen:1978yd}) is not NSVZ \cite{Jack:1996vg,Jack:1996cn,Jack:1998uj,Harlander:2006xq,Mihaila:2013wma}. The MOM scheme is also not NSVZ \cite{Kataev:2014gxa}. Nevertheless, the $\overline{\mbox{DR}}$ calculations implicitly confirm the NSVZ equation, because they demonstrate the validity of scheme-independent consequences following from the NSVZ equation \cite{Kataev:2014gxa,Kataev:2013csa}. These scheme independent relations appear because some terms in the renormalization group functions (RGFs) remain invariant under finite renormalizations, see, e.g., \cite{Kataev:2013vua}. The fact that these relations are satisfied indicates the existence of NSVZ schemes. In fact, there are an infinite number of the NSVZ schemes which constitute a continuous set. In the Abelian case this set has been described in Ref. \cite{Goriachuk:2018cac} and, in particular, includes the on-shell \cite{Kataev:2019olb} and HD+MSL \cite{Kataev:2013eta} schemes. The latter scheme is obtained if a theory is regularized by higher covariant derivatives \cite{Slavnov:1971aw,Slavnov:1972sq} (see Refs. \cite{Krivoshchekov:1978xg,West:1985jx} for  supersymmetric versions of this regularization) and the divergences are removed by minimal subtractions of logarithms, when only powers of $\ln\Lambda/\mu$ are included into renormalization constants \cite{Shakhmanov:2017wji,Stepanyantz:2017sqg}. Equivalently, this scheme can be introduced by imposing certain boundary conditions on the renormalization constants \cite{Kataev:2013eta}. Presumably, the HD+MSL scheme is also NSVZ in the non-Abelian case \cite{Stepanyantz:2016gtk}. This is confirmed by some explicit calculations made in such an approximation where the scheme dependence is essential \cite{Shakhmanov:2017soc,Kazantsev:2018nbl,Kuzmichev:2019ywn}.

The reason why the HD+MSL scheme turns out to be NSVZ is that the NSVZ equation is satisfied by RGFs defined in terms of the bare couplings for theories regularized by higher covariant derivatives. Such RGFs are independent of a renormalization prescription for a fixed regularization, so that Eq. (\ref{NSVZ_Equation_Old}) holds for them in an arbitrary substraction scheme. However, the calculation of Ref. \cite{Aleshin:2016rrr} indicates that with dimensional reduction these RGFs do not satisfy the NSVZ equation starting from the three-loop approximation for the $\beta$-function. For Abelian theories regularized by higher derivatives the validity of the NSVZ equation for RGFs defined in terms of the bare couplings has been proved in all loops in Refs. \cite{Stepanyantz:2011jy,Stepanyantz:2014ima}. This proof is based on the fact that the loop integrals which give the $\beta$-function defined in terms of the bare coupling constant in supersymmetric theories are integrals of double total derivatives with respect to the loop momenta. The factorization into total and double total derivatives has first been noted in calculating the lowest-order quantum corrections for ${\cal N}=1$ supersymmetric electrodynamics (SQED) in Refs. \cite{Soloshenko:2003nc} and \cite{Smilga:2004zr}, respectively. The subsequent generalizations of the proof made in Ref. \cite{Stepanyantz:2011jy} allowed deriving the all-loop NSVZ-like equations for the Adler $D$-function \cite{Adler:1974gd} in ${\cal N}=1$ supersymmetric chromodynamics \cite{Shifman:2014cya,Shifman:2015doa} and for the renormalization of the photino mass in softly broken ${\cal N}=1$ SQED \cite{Nartsev:2016nym}. (In softly broken supersymmetric theories the NSVZ-like equation for the renormalization of the gaugino mass has first been found in \cite{Hisano:1997ua,Jack:1997pa,Avdeev:1997vx}.) In both cases the HD+MSL scheme is NSVZ \cite{Kataev:2017qvk,Nartsev:2016mvn}, because in this scheme RGFs defined in terms of the renormalized couplings coincide with RGFs defined in terms of the bare couplings up to the renaming of arguments, see Eq. (\ref{HD+MSL_RGF}) below.

In the non-Abelian case the all-loop proof of the NSVZ equation by a similar method has not yet been completed, although its main steps are at present quite clear. First, the NSVZ equation should be rewritten in the equivalent form \cite{Stepanyantz:2016gtk}

\begin{equation}\label{NSVZ_Equation_New}
\frac{\beta(\alpha,\lambda)}{\alpha^2} = -\frac{1}{2\pi} \Big(3C_2 - T(R) - 2C_2\gamma_c(\alpha,\lambda) - 2C_2\gamma_V(\alpha,\lambda) + C(R)_i{}^j(\gamma_{\phi})_j{}^i(\alpha,\lambda)/r \Big)
\end{equation}

\noindent
with the help of Eq. (\ref{Z_Triple_Relation}), where $\gamma_c$ and $\gamma_V$ are the anomalous dimensions of the Faddeev--Popov ghosts and the quantum gauge superfield, respectively. According to \cite{Stepanyantz:2019ihw} the $\beta$-function is given by integrals of double total derivatives with respect to the loop momenta in all loops if the higher derivatives are used for regularization. (Again, in the case of using dimensional reduction it is not so \cite{Aleshin:2015qqc}.) Certainly, this is confirmed by a large number of explicit calculations \cite{Shakhmanov:2017soc,Kazantsev:2018nbl,Pimenov:2009hv,Stepanyantz:2011bz,Aleshin:2016yvj}. Due to this structure of the loop integrals the $\beta$-function beyond the one-loop approximation is given by a sum of $\delta$-singularities, which appear when double total derivatives act on an inverse squared momentum. The all-loop sums of singularities which occur when the double total derivatives act on inverse squared momenta of the matter superfields and of the Faddeev--Popov ghosts give the corresponding anomalous dimensions in Eq. (\ref{NSVZ_Equation_New}) \cite{Stepanyantz:2019lfm}. For remaining singularities produced by momenta of the quantum gauge superfield the corresponding paper is in preparation. Thus, there are strong indications that the NSVZ equation is satisfied by RGFs defined in terms of the bare couplings for theories regularized by higher covariant derivatives independently of a way of renormalization.

The above discussion demonstrates that the higher covariant derivative regularization helps to reveal the underlying structure of quantum corrections which is responsible for the appearance of the NSVZ equation in perturbation theory. That is why it is especially interesting to use it for investigating ${\cal N}=1$ finite theories (see Ref. \cite{Heinemeyer:2019vbc} for a recent review of the theoretical aspects and the phenomenological applications). The direct calculation of Ref. \cite{Parkes:1984dh} made in the  $\overline{\mbox{DR}}$ scheme demonstrated that if an ${\cal N}=1$ supersymmetric gauge theory is finite in the one-loop approximation, then it is also finite in the two-loop approximation. The same result was obtained from arguments based on anomalies \cite{Jones:1983vk,Jones:1984cx}. According to \cite{Grisaru:1985tc}, for theories finite in the $L$-th loop the $\beta$-function vanishes in the $(L+1)$-th loop. The same statement immediately follows from the NSVZ equation. We know that the NSVZ equation naturally appears with the higher covariant derivative regularization and is not valid in the $\overline{\mbox{DR}}$ scheme. Therefore, we are tempted to suggest that the higher derivative regularization could reveal some features of one-loop finite ${\cal N}=1$ supersymmetric theories leading to their possible all-loop finiteness. For this purpose one should investigate the anomalous dimension of the matter superfields. The calculation made in Ref. \cite{Parkes:1985hh} demonstrated that in the $\overline{\mbox{DR}}$ scheme the three-loop anomalous dimension does not vanish. However, it has explicitly been verified \cite{Jack:1996qq} that it is possible to tune a subtraction scheme in such a way that a one-loop finite theory will also be finite in the three-loop approximation. According to the general argumentation of Refs. \cite{Kazakov:1986bs,Ermushev:1986cu,Lucchesi:1987he,Lucchesi:1987ef}, a scheme in which a one-loop finite theory is finite in all loops should exist, but at present there is no simple prescription for constructing it. Possibly, the use of the higher covariant derivative regularization could help to solve this problem.

One more interesting subject is the possible existence of the exact expression for the anomalous dimension of the matter superfields  for theories obeying the $P=\frac{1}{3}Q$ constraint proposed by Jack and Jones in Ref. \cite{Jack:1995gm}. In our notation it can be written as

\begin{equation}\label{JJN_Constraint}
\lambda^*_{imn}\lambda^{jmn} - 4\pi\alpha C(R)_i{}^j = \frac{2\pi\alpha}{3} Q \delta_i{}^j,\qquad \mbox{where}\qquad Q \equiv T(R) - 3C_2.
\end{equation}

\noindent
According to Ref. \cite{Jack:1996qq}, for theories which satisfy the condition (\ref{JJN_Constraint}) the anomalous dimension of the matter superfields can possibly be written in the Jack, Jones, North (JJN) form

\begin{equation}\label{Gamma_Exact}
(\gamma_\phi)_i{}^j(\alpha,\lambda)\ \to\ (\gamma_\phi)_i{}^j(\alpha) = \frac{\alpha Q}{6\pi (1+\alpha Q/6\pi)}\delta_i^j,
\end{equation}

\noindent
while the $\beta$-function does not depend on the Yukawa couplings and is given by the geometric series

\begin{equation}\label{Beta_JJN_Exact}
\beta(\alpha,\lambda)\ \to\ \beta(\alpha) = \frac{\alpha^2 Q}{2\pi (1+\alpha Q/6\pi)}.
\end{equation}

\noindent
(Note that here we again do not specify a definition of RGFs, a regularization, and a renormalization prescription.) Although the three-loop calculation of Ref. \cite{Jack:1996qq} presumably excludes this possibility, this particular case seems to be very interesting and deserving a further investigation. As a justification, it is possible to suggest that Eq. (\ref{Gamma_Exact}) can be valid in higher loops if some more constraints are imposed on the theory together with Eq. (\ref{JJN_Constraint}).

In this paper we consider a general renormalizable ${\cal N}=1$ supersymmetric theory regularized by higher covariant derivatives, which is described in Sect. \ref{Section_Theory_Considered}. In Sect. \ref{Section_Two_Lool_Gamma_Bare} for this theory we calculate the two-loop anomalous dimension of the matter superfields defined in terms of the bare couplings. It is demonstrated that there is a simple relation between the higher derivative regulators and the Pauli--Villars masses for which it vanishes for the one-loop finite theories. Also there is a regularization for which Eq. (\ref{Gamma_Exact}) is valid for the anomalous dimension defined in terms of the bare couplings under the $P=\frac{1}{3}Q$ constraint. Certainly, the same statements are valid for the anomalous dimension defined in terms of the renormalized couplings in the HD+MSL scheme. For a general renormalization prescription the expression for the anomalous dimension defined in terms of the renormalized couplings is found in Sect. \ref{Section_Two_Loop_Gamma_Renormalized}. Using the statement that the NSVZ equation is presumably valid for RGFs defined in terms of the bare couplings with the higher derivative regularization, the expression for the three-loop $\beta$-function is written in Sect. \ref{Section_Three_Loop_Beta}. Again this is done for the $\beta$-functions defined both in terms of the bare couplings and in terms of the renormalized couplings. The particular cases of the one-loop finite theories and theories satisfying the constraint (\ref{JJN_Constraint}) are investigated. Also in Sect. \ref{Section_Three_Loop_Beta} we demonstrate that for the one-loop finite theories the NSVZ equation in the considered approximation is valid in an arbitrary subtraction scheme.

\section{The theory under consideration}
\hspace*{\parindent}\label{Section_Theory_Considered}

We consider the ${\cal N}=1$ SYM theory with a simple gauge group $G$ interacting with the chiral matter superfields $\phi_i$ in a representation $R$, which can in general be reducible. At the classical level this theory in the massless limit is described by the superfield action \cite{Gates:1983nr,West:1990tg,Buchbinder:1998qv}

\begin{equation}\label{Action_Massless_Classical}
S = \frac{1}{2e_0^2} \mbox{tr Re} \int d^4x\, d^2\theta\, W^{a} W_{a} + \frac{1}{4} \int d^4x\, d^4\theta\, \phi^{* i}  \big(e^{2V}\big)_i{}^j \phi_j + \Big( \frac{1}{6}\int d^4x\, d^2\theta\, \lambda_0^{ijk} \phi_{i} \phi_{j} \phi_{k} + \mbox{c.c.}\Big),
\end{equation}

\noindent where the supersymmetric gauge field strength is given by the chiral superfield $W_a = \bar D^2\big(e^{-2V} D_a e^{2V}\big)/8$. Note that $V=e_0V^{A}t^{A}$ in the first term of the action (\ref{Action_Massless_Classical}), while $V=e_0V^{A}T^A$ in the second one.

To quantize the theory, one should take into account that the quantum gauge superfield is renormalized in a nonlinear way \cite{Piguet:1981fb,Piguet:1981hh,Tyutin:1983rg}. Also it is convenient to use the background field method \cite{DeWitt:1965jb,Abbott:1980hw,Abbott:1981ke} in the superfield formulation \cite{Grisaru:1982zh,Gates:1983nr}, because it produces a manifestly gauge invariant effective action. All this can be achieved by making the replacement

\begin{equation}\label{Replacement}
e^{2V} \to e^{2{\cal F}(V)} e^{2\bm{V}},
\end{equation}

\noindent
where $\bm{V}$ and $V$ are the background and quantum gauge superfields, respectively. Note that in our notation the latter superfield satisfies the constraint $V^+ = e^{-2\bm{V}} V e^{2\bm{V}}$. The parameters of the nonlinear renormalization are included into the function

\begin{equation}\label{Nonlinear_Function}
{\cal F}(V) = V+ O(V^3).
\end{equation}

\noindent
The first nonlinear term in this function has been calculated in Refs. \cite{Juer:1982fb,Juer:1982mp}. Subsequently it was demonstrated that the presence of the nonlinear renormalization is very essential for the renormalization group equations to be satisfied \cite{Kazantsev:2018kjx}.

After the replacement (\ref{Replacement}), the action regularized by higher covariant derivatives in the massless limit can be written as

\begin{eqnarray}\label{Regularized_Action_Massless_Classical}
&& S_{\mbox{\scriptsize reg}} = \frac{1}{2 e_0^2}\mbox{Re}\, \mbox{tr} \int d^4x\,d^2\theta\, W^a \Big[e^{-2\bm{V}} e^{-2{\cal F}(V)}\,  R\Big(-\frac{\bar\nabla^2 \nabla^2}{16\Lambda^2}\Big)\, e^{2{\cal F}(V)}e^{2\bm{V}}\Big]_{\mbox{\scriptsize Adj}} W_a \nonumber\\
&& + \frac{1}{4} \int d^4x\,d^4\theta\, \phi^{*i} \Big[\, F\Big(-\frac{\bar\nabla^2 \nabla^2}{16\Lambda^2}\Big) e^{2{\cal F}(V)}e^{2\bm{V}}\Big]_i{}^j \phi_j
+ \Big(\frac{1}{6} \lambda_0^{ijk} \int d^4x\, d^2\theta\, \phi_i \phi_j \phi_k + \mbox{c.c.} \Big),\qquad
\end{eqnarray}

\noindent
where the left and right covariant spinor derivatives are given by the equations

\begin{equation}
\bar\nabla_{\dot a} \equiv e^{2{\cal F}(V)} e^{2\bm{V}} \bar D_{\dot a} e^{-2\bm{V}} e^{-2{\cal F}(V)}; \qquad \nabla_a \equiv D_a,
\end{equation}

\noindent
respectively. The regulator functions $F(x)$ and $R(x)$ are infinite at infinity and approach 1 at $x=0$. Note that in Eq. (\ref{Regularized_Action_Massless_Classical}) the gauge superfield strength is defined as

\begin{equation}
W_a \equiv \frac{1}{8} \bar D^2 \Big[e^{-2\bm{V}} e^{-2{\cal F}(V)}\, D_a \Big(e^{2{\cal F}(V)}e^{2\bm{V}}\Big)\Big].
\end{equation}

In this paper we will use the gauge fixing term

\begin{equation}\label{Gauge_Fixing_Action}
S_{\mbox{\scriptsize gf}} = - \frac{1}{16\xi_0 e_0^2}\, \mbox{tr} \int d^4x\,d^4\theta\,  \bm{\nabla}^2 V  R\Big(-\frac{\bm{\bar\nabla}^2 \bm{\nabla}^2}{16\Lambda^2}\Big)_{\mbox{\scriptsize Adj}} \bm{\bar\nabla}^2 V
\end{equation}

\noindent containing the parameter $\xi_0$. Due to the presence of the background covariant derivatives

\begin{equation}
\bm{\bar\nabla}_{\dot a} \equiv e^{2\bm{V}} \bar D_{\dot a} e^{-2\bm{V}};\qquad \bm{\nabla}_a \equiv D_a
\end{equation}

\noindent
it is invariant under the background gauge symmetry. According to \cite{Aleshin:2016yvj} the one-loop renormalization of the gauge parameter is described by the equation

\begin{equation}
\frac{1}{\xi_0e_0^2} = \frac{1}{\xi e^2} + \frac{C_2(1-\xi)}{12\pi^2\xi} \Big(\ln\frac{\Lambda}{\mu} + a_1\Big) + O(e^2),
\end{equation}

\noindent
where $a_1$ is a finite constant which originates from the arbitrariness in choosing a renormalization prescription. In this paper we will use the Feynman gauge $\xi=1$ in which

\begin{equation}
\xi_0 e_0^2 = e^2 + O(e^6).
\end{equation}

\noindent
The quantization procedure (see, e.g., \cite{West:1990tg}) also requires introducing the Faddeev--Popov and Nielsen--Kallosh ghosts. Their actions ($S_{\mbox{\scriptsize FP}}$ and $S_{\mbox{\scriptsize NK}}$, respectively) can be found, e.g., in Ref.~\cite{Stepanyantz:2019lyo}.

Due to the presence of the higher derivative regulators $R(x)$ and $F(x)$ in the actions (\ref{Regularized_Action_Massless_Classical}) and (\ref{Gauge_Fixing_Action}) all divergences disappear beyond the one-loop approximation. This is a general feature of the higher covariant derivative regularization, see, e.g., \cite{Faddeev:1980be}. For regularizing the remaining one-loop divergences one has to introduce the Pauli--Villars determinants \cite{Slavnov:1977zf}. Following Refs. \cite{Aleshin:2016yvj,Kazantsev:2017fdc}, we define the generating functional as

\begin{equation}\label{Z_Functional}
Z[\mbox{Sources}] = \int D\mu\; \mbox{Det}(PV, M_{\varphi})^{-1} \big(\mbox{Det}(PV, M)\big)^c\, \exp\Big\{i\Big(S_{\mbox{\scriptsize reg}} + S_{\mbox{\scriptsize gf}} + S_{\mbox{\scriptsize FP}} + S_{\mbox{\scriptsize NK}} + S_{\mbox{\scriptsize sources}}\Big)\Big\},
\end{equation}

\noindent where the Pauli--Villars determinants are given by the functional integrals

\begin{equation}
\mbox{Det}(PV, M_{\varphi})^{-1} = \int D\varphi_1 D\varphi_2 D\varphi_3\, e^{iS_{\varphi}}; \qquad \mbox{Det}(PV, M)^{-1} = \int D\Phi\, e^{iS_\Phi}.
\end{equation}

\noindent
Here $\varphi_a$ is a set of three commuting chiral superfields in the adjoint representation with the action

\begin{eqnarray}
&& S_{\varphi}=\frac{1}{2e_0^2}\mbox{tr}\int d^4x\, d^4\theta\, \Big\{\varphi_{1}^{+} \Big[ R\Big(-\frac{\bar\nabla^2{\nabla}^2}{16{\Lambda}^2}\Big) e^{2{\cal F}(V)} e^{2\bm{V}}\Big]_{\mbox{\scriptsize Adj}}\varphi_{1} + \varphi_{2}^{+}\big[e^{2{\cal F}(V)} e^{2\bm{V}}\big]_{\mbox{\scriptsize Adj}}\varphi_{2} \qquad\nonumber\\
&& + \varphi_{3}^{+} \left[e^{2{\cal F}(V)} e^{2\bm{V}}\right]_{\mbox{\scriptsize Adj}}\varphi_{3}\Big\} + \frac{1}{2e_0^2}\Big(\mbox{tr}\int d^4x\, d^2\theta\, M_{\varphi}(\varphi_{1}^2+\varphi_{2}^2 + \varphi_{3}^2) + \mbox{c.c.}\Big),
\end{eqnarray}

\noindent
and $\Phi_i$ is a multiplet of commuting chiral superfields in a representation $R_{PV}$ that admits a gauge invariant mass term with the action

\begin{equation}
S_{\Phi} = \frac{1}{4} \int d^4x\, d^4\theta\, \Phi^{*i}\Big[ F\Big(-\frac{\overline{\nabla}^2\nabla^2}{16\Lambda^2}\Big) e^{2{\cal F}(V)} e^{2\bm{V}}\Big]_i \Big.^j \Phi_j + \frac{1}{4}\Big(\int d^4x\, d^2\theta\, M^{ij} \Phi_i\Phi_j + \mbox{c.c.}\Big).
\end{equation}

\noindent
We assume that the invariant tensor $M^{ij}$ satisfies the condition

\begin{equation}
M^{ik} M^*_{kj}=M^2\delta^i_j,
\end{equation}

\noindent
and the Pauli--Villars masses are proportional to the constant $\Lambda$ in the higher derivative terms,

\begin{equation}
M_{\varphi}=a_{\varphi}\Lambda;\qquad M=a\Lambda,
\end{equation}

\noindent
where $a_\varphi$ and $a$ are independent of couplings. Then the one-loop divergences and subdivergences will be regularized if the constant $c$ in the generating functional (\ref{Z_Functional}) is equal to $T(R)/T(R_{PV})$.

In this paper we will calculate the anomalous dimension of the matter superfields. To construct it, first, one should consider the part of the effective action which corresponds to self-energy diagrams for the chiral matter superfields. It can written in the form

\begin{equation}\label{Effective}
\Gamma_{\phi}^{(2)} = \frac{1}{4}\int \frac{d^4q}{(2\pi)^4} d^4\theta\, \phi^{*i}(-q,\theta) \phi_j(q,\theta)\, (G_\phi)_i{}^j(\alpha_0,\lambda_0,q^2/\Lambda^2),
\end{equation}

\noindent where $(G_\phi)_i{}^j = \delta_i^j + (\Delta G_\phi)_i{}^j$, and $(\Delta G_\phi)_i{}^j$ is of order $\alpha_0$ or $\lambda_0^2$.

Let $\alpha\equiv e^2/4\pi$ and $\lambda^{ijk}$ denote the renormalized gauge and Yukawa coupling constants, respectively. The bare and renormalized chiral matter superfields are related by the equation

\begin{equation}
\phi_i = (\sqrt{Z_\phi})_i{}^j(\phi_R)_j,
\end{equation}

\noindent where the renormalization constant $(Z_\phi)_i{}^j$ is determined by requiring the finiteness of the product $(Z_\phi)_i{}^k (G_\phi)_k{}^j$ expressed in terms of the renormalized couplings in the limit $\Lambda\to \infty$. Note that this requirement does not allow fixing this constant uniquely. The remaining arbitrariness is removed by choosing a renormalization prescription.

Divergences of a theory are encoded in RGFs. According to \cite{Kataev:2013eta}, it is necessary to distinguish between RGFs defined in terms of the bare couplings and the ones (standardly) defined in terms of the renormalized couplings. In terms of the bare couplings, the definitions of the $\beta$-function and the anomalous dimension of the chiral matter superfields read as

\begin{eqnarray}\label{Beta_Definition_Bare}
&& \beta(\alpha_0,\lambda_0) = \frac{d\alpha_0}{d\ln\Lambda}\bigg|_{\alpha,\lambda=\mbox{\scriptsize const}}; \qquad\\
\label{Gamma_Definition_Bare}
&& (\gamma_\phi)_i{}^j(\alpha_0,\lambda_0) = -\frac{d(\ln Z_\phi)_i{}^j}{d\ln\Lambda} \bigg|_{\alpha,\lambda=\mbox{\scriptsize const}}.
\end{eqnarray}

\noindent
For a fixed regularization these RGFs are independent of a renormalization procedure. That is why they (presumably) satisfy the NSVZ relation in an arbitrary subtraction scheme supplementing the higher covariant derivative regularization. In terms of the renormalized couplings RGFs are defined by the equations

\begin{eqnarray}\label{Beta_Definition_Renormalized}
&& \widetilde \beta(\alpha,\lambda) = \frac{d\alpha}{d\ln\mu}\bigg|_{\alpha_0,\lambda_0=\mbox{\scriptsize const}}; \qquad\\
\label{Gamma_Definition_Renormalized}
&& (\widetilde{\gamma}_\phi)_i{}^j(\alpha,\lambda) = \frac{d(\ln Z_\phi)_i{}^j}{d\ln\mu} \bigg|_{\alpha_0,\lambda_0=\mbox{\scriptsize const}},
\end{eqnarray}

\noindent
where the derivatives are taken with respect to the logarithm of the renormalization point $\mu$ at fixed values of the bare couplings. The $\beta$-function (\ref{Beta_Definition_Renormalized}) and the anomalous dimension (\ref{Gamma_Definition_Renormalized}) are scheme-dependent starting from the three- and two-loop approximation, respectively. Up to the renaming of arguments both definitions of RGFs coincide in the HD+MSL scheme, in which the renormalization constants contain only powers of $\ln\Lambda/\mu$ and all finite constants fixing a subtraction scheme are set to 0,

\begin{equation}\label{HD+MSL_RGF}
\beta(\alpha_0\to \alpha,\lambda_0\to\lambda) = \widetilde\beta(\alpha,\lambda)\Big|_{\mbox{\scriptsize HD+MSL}}; \qquad (\gamma_\phi)_i{}^j(\alpha_0\to \alpha,\lambda_0\to\lambda) = (\widetilde\gamma_\phi)_i{}^j(\alpha,\lambda)\Big|_{\mbox{\scriptsize HD+MSL}}.
\end{equation}

RGFs defined in terms of the bare couplings are related to the corresponding Green functions. For example, using the finiteness of the expression $(Z_\phi G_\phi)_i{}^j$ the anomalous dimension (\ref{Gamma_Definition_Bare}), which will be calculated in this paper, can be presented in the equivalent form

\begin{equation}\label{Gamma_From_G}
(\gamma_\phi)_i{}^j(\alpha_0,\lambda_0) = \frac{d(\ln G_\phi)_i{}^j}{d\ln\Lambda}\biggr|_{\alpha,\lambda=\mbox{\scriptsize const};\, q\to 0},
\end{equation}

\noindent where the limit $q\to 0$ is necessary here to get rid of the terms vanishing in the limit $\Lambda\to\infty$.

To use Eq. (\ref{Gamma_From_G}), one needs to know how the renormalized and bare couplings are related at lower orders. In particular, for calculating the two-loop anomalous dimension of the matter superfields, one needs the one-loop relation between the bare and renormalized couplings. Moreover, we will also need the two-loop relation between the bare and renormalized gauge coupling constants for investigating the three-loop $\beta$-function defined in terms of the renormalized couplings in Sect. \ref{Subsection_Three_Loop_Beta_Renormalized}. It can be found by integrating Eq. (\ref{Beta_Definition_Bare}) in which we substitute the two-loop $\beta$-function. With the regularization considered in this paper it has been calculated in Ref. \cite{Stepanyantz:2019lyo}. The result contains some arbitrary integration constants $b_i$, which reflect the arbitrariness in choosing a subtraction scheme,

\begin{eqnarray}\label{Two_Loop_Alpha}
&&\hspace*{-7mm} \frac{1}{\alpha} - \frac{1}{\alpha_0} = -\frac{3}{2\pi}C_2\Big(\ln \frac{\Lambda}{\mu} + b_{11}\Big) + \frac{1}{2\pi} T(R) \Big(\ln \frac{\Lambda}{\mu} + b_{12}\Big) - \frac{3\alpha}{4\pi^2} C_2^2 \Big(\ln\frac{\Lambda}{\mu} + b_{21} \Big) + \frac{\alpha}{4\pi^2 r}  C_2\nonumber\\
&&\hspace*{-7mm} \times \mbox{tr}\, C(R) \Big(\ln \frac{\Lambda}{\mu} + b_{22}\Big) + \frac{\alpha}{2\pi^2 r} \mbox{tr}\left[C(R)^2\right] \Big(\ln \frac{\Lambda}{\mu} + b_{23}\Big) - \frac{1}{8\pi^3 r} C(R)_j{}^i \lambda^*_{imn} \lambda^{jmn} \Big(\ln \frac{\Lambda}{\mu}  + b_{24}\Big)\nonumber\\
&&\hspace*{-7mm} + O(\alpha^2,\alpha \lambda^2,\lambda^4). \vphantom{\frac{1}{2}}
\end{eqnarray}

\noindent
Due to the non-renormalization theorem of Ref. \cite{Grisaru:1979wc}, the superpotential does not receive divergent radiative corrections. This implies that the renormalization of the Yukawa couplings is related to the renormalization of the matter superfields by the equation

\begin{equation}\label{Yukawa_Renormalization}
\lambda^{ijk}=(\sqrt{Z_\phi})_l{}^{i}(\sqrt{Z_\phi})_m{}^j(\sqrt{Z_\phi})_n{}^k\lambda_0^{lmn}.
\end{equation}

\noindent
In this paper we will consider only such renormalization schemes for which it is valid (although, in general, it is possible to construct subtraction schemes breaking this equation). The one-loop expression for the renormalization constant $(Z_\phi)_i{}^j$ can be written as

\begin{equation}\label{One_Loop_Z}
(Z_\phi)_i{}^j(\alpha,\lambda) = \delta_i{}^j + \frac{\alpha}{\pi} C(R)_i{}^j \Big(\ln\frac{\Lambda}{\mu}+g_{11}\Big) - \frac{1}{4\pi^2} \lambda^*_{imn}\lambda^{jmn} \Big(\ln\frac{\Lambda}{\mu} + g_{12} \Big)
+ O(\alpha^2,\alpha\lambda^2,\lambda^4),
\end{equation}

\noindent where $g_{11}$ and $g_{12}$ are arbitrary constants similar to $b_i$ in Eq. (\ref{Two_Loop_Alpha}). Eqs. (\ref{Yukawa_Renormalization}) and (\ref{One_Loop_Z}) determine the one-loop renormalization of the Yukawa couplings,

\begin{eqnarray}\label{One_Loop_Lambda}
&&\hspace*{-5mm} \lambda_0^{ijk} = \lambda^{ijk} - \frac{\alpha}{2\pi} \Big(C(R)_m{}^i \lambda^{mjk} + C(R)_m{}^j \lambda^{imk} + C(R)_m{}^k \lambda^{ijm} \Big) \Big(\ln\frac{\Lambda}{\mu} + g_{11}\Big) + \frac{1}{8\pi^2} \Big( \lambda^{mjk} \nonumber\\
&&\hspace*{-5mm} \times \lambda^*_{mab}\lambda^{iab} + \lambda^{imk}\lambda^*_{mab}\lambda^{jab} + \lambda^{ijm} \lambda^*_{mab}\lambda^{kab} \Big) \Big(\ln\frac{\Lambda}{\mu} + g_{12}\Big)
+ O\Big(\alpha^2\lambda,\alpha\lambda^3,\lambda^5\Big).
\end{eqnarray}

\noindent
By definition, in the HD+MSL renormalization scheme all constants $b_i$ and $g_i$ are set to 0.

\section{Two-loop anomalous dimension}
\hspace*{\parindent}\label{Section_Two_Lool_Gamma_Bare}

\begin{figure}[h]
\hspace*{1mm}
\begin{picture}(0,15)
\put(0.5,14.0){$(1)$}
\put(0.5,12.6){\includegraphics[scale=0.45]{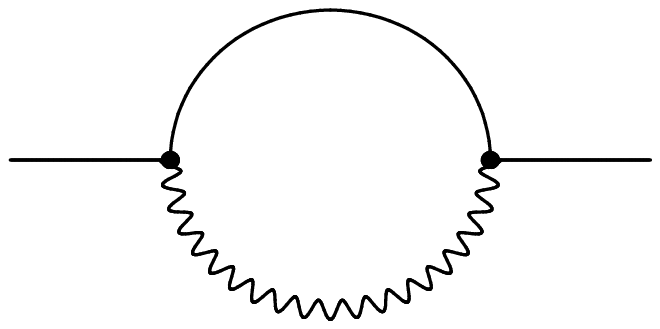}}
\put(4.5,14.0){$(2)$}
\put(4.5,12.6){\includegraphics[scale=0.45]{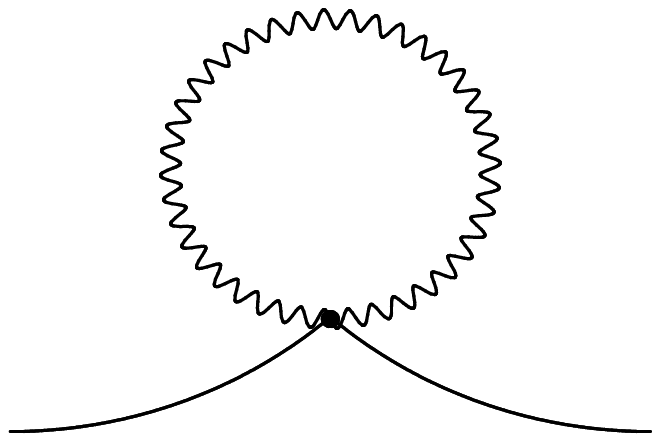}}
\put(8.5,14.0){$(3)$}
\put(8.5,12.6){\includegraphics[scale=0.45]{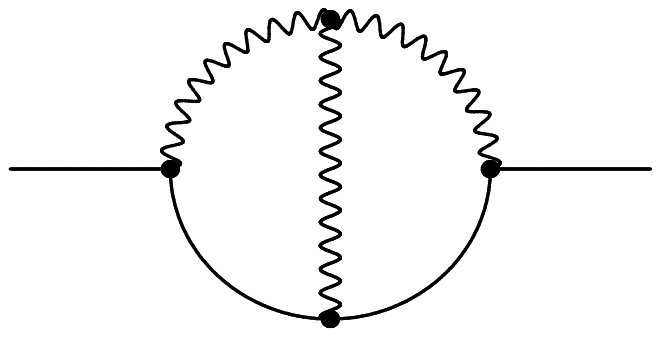}}
\put(12.5,14.0){$(4)$}
\put(12.5,12.6){\includegraphics[scale=0.45]{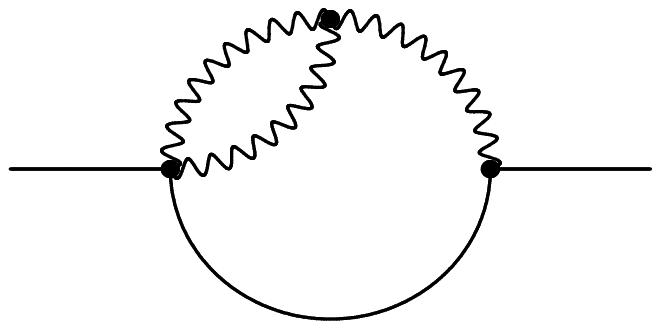}}
\put(0.5,11.0){$(5)$}
\put(0.5, 9.6){\includegraphics[scale=0.45]{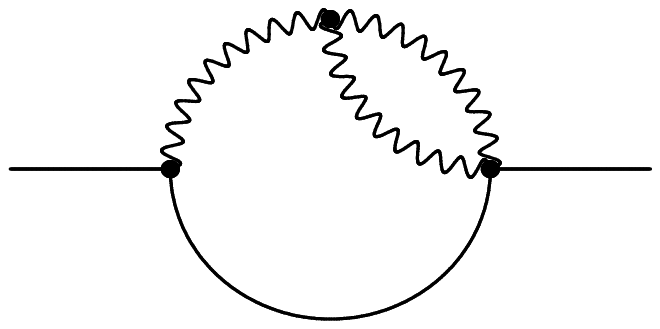}}
\put(4.5,11.0){$(6)$}
\put(4.5, 9.6){\includegraphics[scale=0.45]{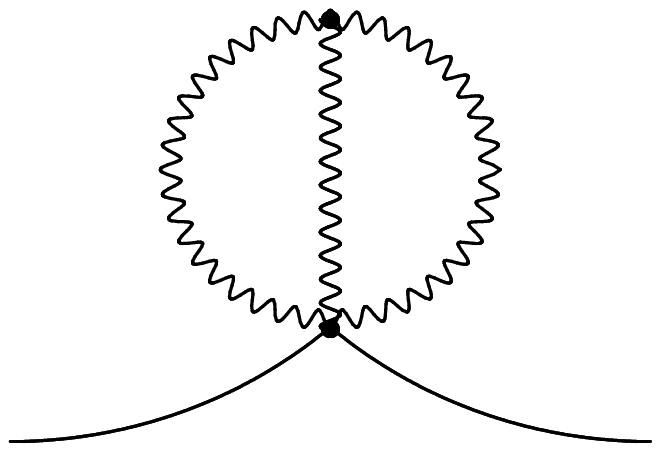}}
\put(8.5,11.0){$(7)$}
\put(8.5, 9.6){\includegraphics[scale=0.45]{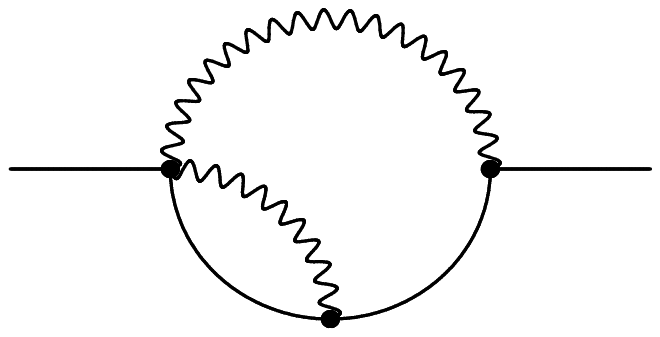}}
\put(12.5,11.0){$(8)$}
\put(12.5, 9.6){\includegraphics[scale=0.45]{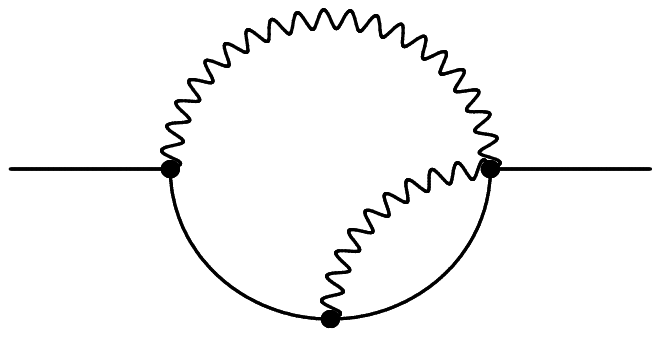}}
\put(0.5,8.0){$(9)$}
\put(0.3, 6.6){\includegraphics[scale=0.4]{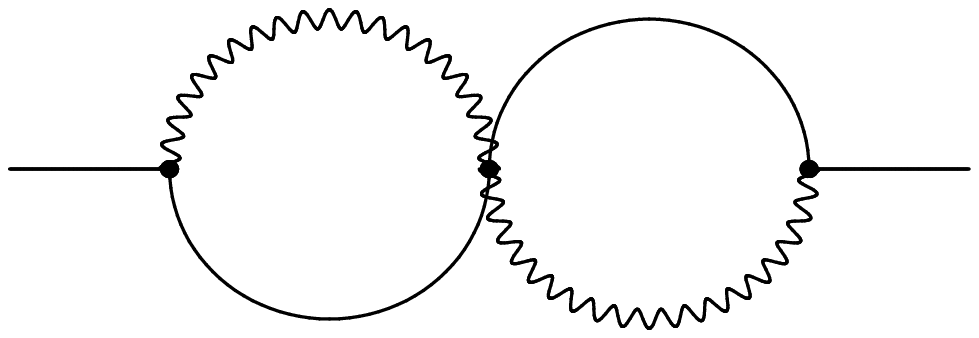}}
\put(4.5,8.0){$(10)$}
\put(4.5, 6.6){\includegraphics[scale=0.45]{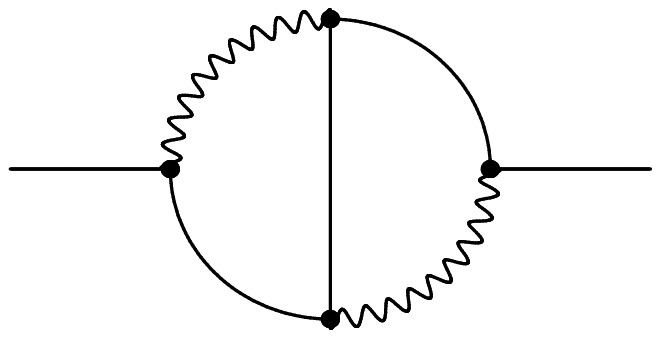}}
\put(8.5,8.0){$(11)$}
\put(8.5, 6.6){\includegraphics[scale=0.45]{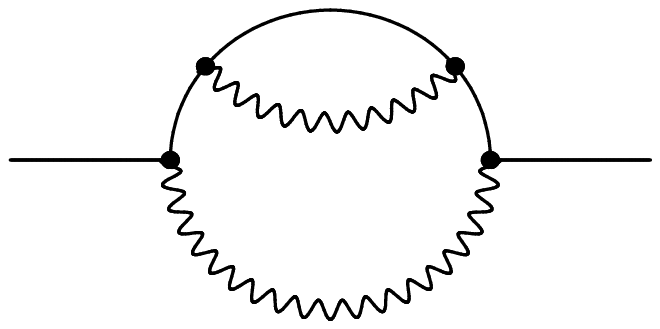}}
\put(12.5,8.0){$(12)$}
\put(12.5, 6.6){\includegraphics[scale=0.45]{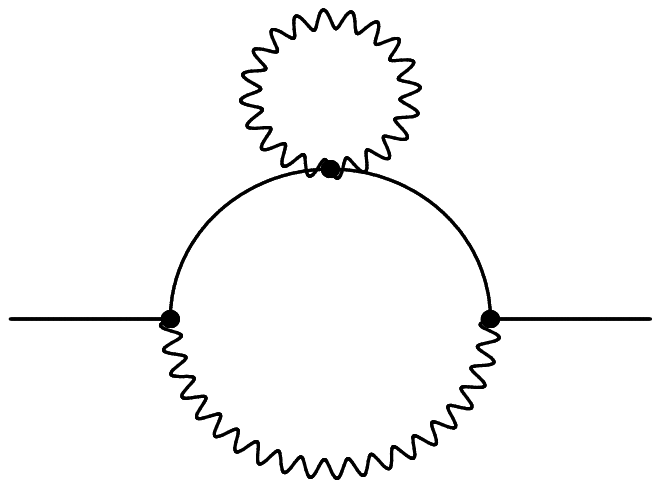}}
\put(0.5,5.0){$(13)$}
\put(0.5, 3.6){\includegraphics[scale=0.45]{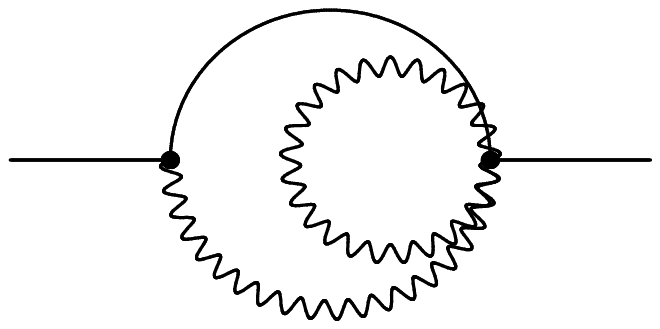}}
\put(4.5,5.0){$(14)$}
\put(4.5, 3.6){\includegraphics[scale=0.45]{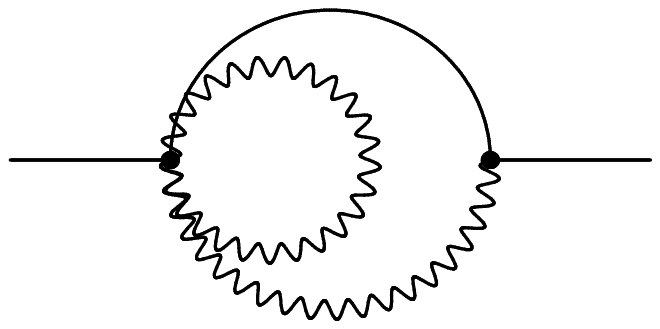}}
\put(8.5,5.0){$(15)$}
\put(8.5, 3.6){\includegraphics[scale=0.45]{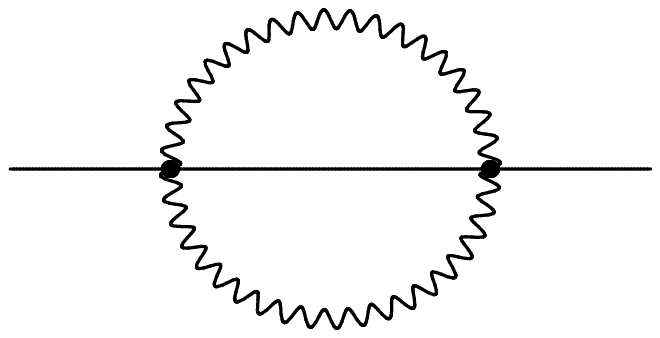}}
\put(12.5,5.0){$(16)$}
\put(12.8, 3.3){\includegraphics[scale=0.45]{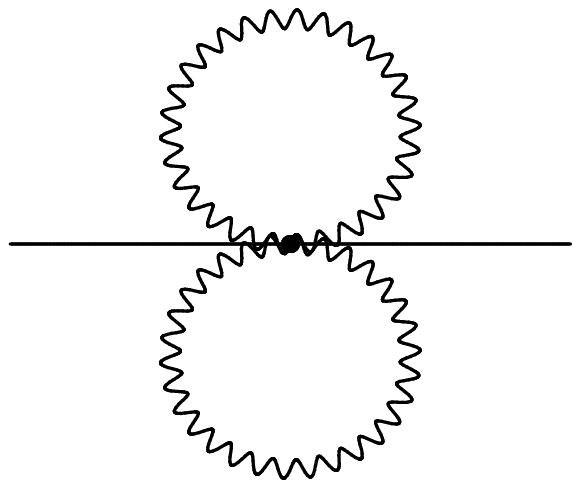}}
\put(4.5,2.0){$(17)$}
\put(4.5, 0.6){\includegraphics[scale=0.45]{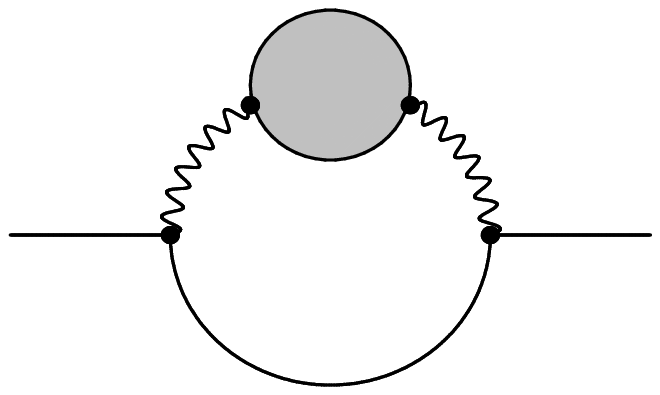}}
\put(8.5,2.0){$(18)$}
\put(8.5, 0.6){\includegraphics[scale=0.45]{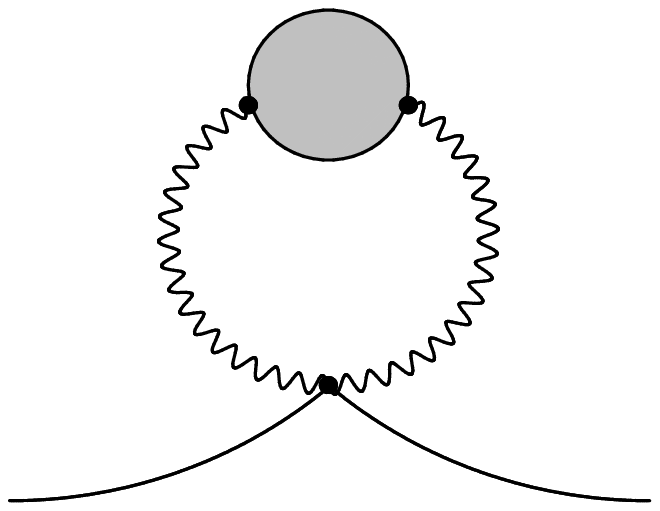}}
\end{picture}
\caption{Superdiagrams without Yukawa vertices contributing to the two-loop anomalous dimension of the matter superfields. In the diagrams (17) and (18) the gray circles denote insertions of the one-loop polarization operator of the quantum gauge superfield.}
\label{Figure_Diagrams_Without_Yukawa}
\end{figure}

\begin{figure}[h]
\hspace*{5mm}
\begin{picture}(0,5)
\put(0,2){\includegraphics[scale=0.4]{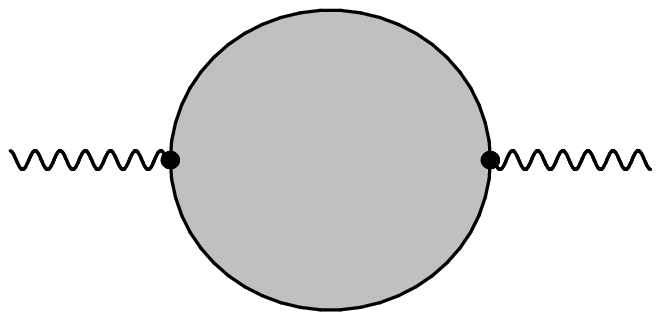}}
\put(3,2.5){=}
\put(4,2.5){$\left\{\vphantom{\begin{array}{c}1\\0\\0\\0\\0\\0\\0\\0\end{array}}\right.$}
\put(5,3.25){\includegraphics[scale=0.4]{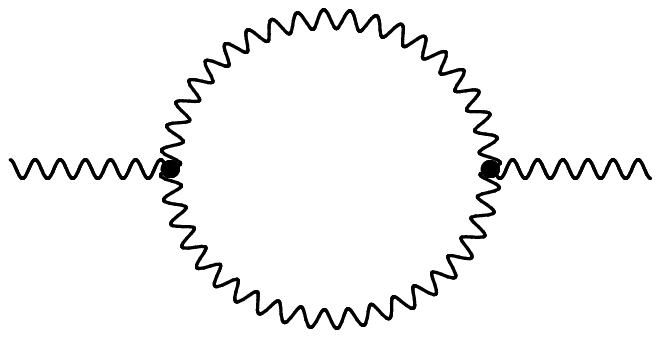}}
\put(5,0.75){\includegraphics[scale=0.4]{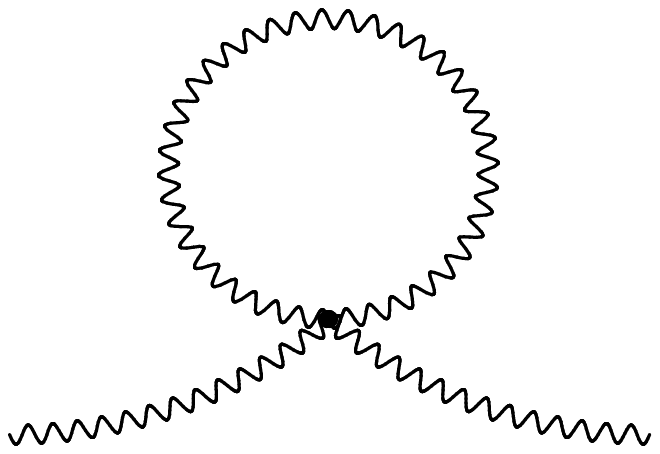}}
\put(8.5,3.25){\includegraphics[scale=0.4]{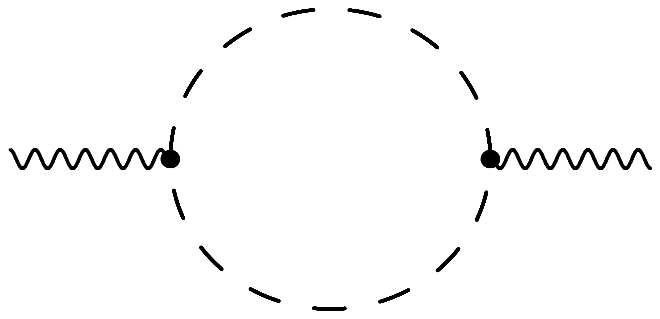}}
\put(8.5,0.75){\includegraphics[scale=0.4]{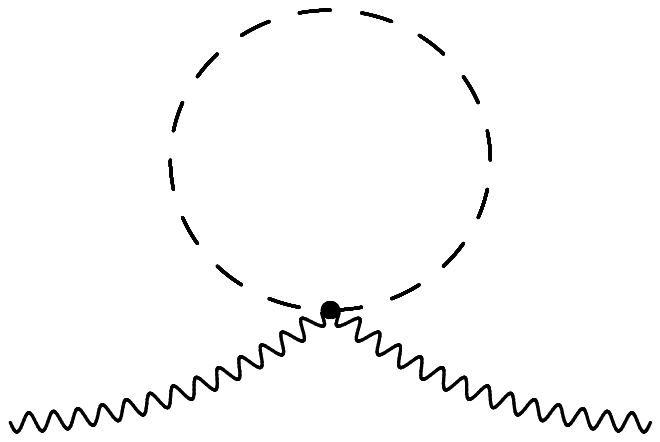}}
\put(12,3.25){\includegraphics[scale=0.4]{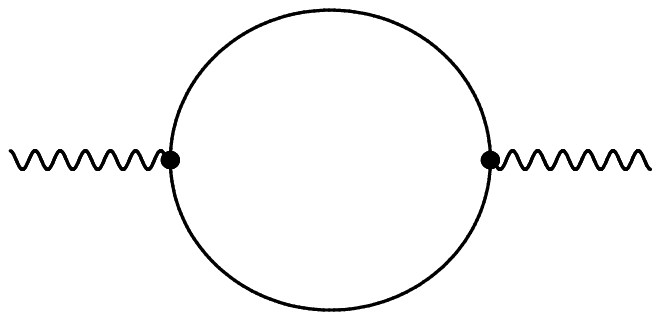}}
\put(12,0.75){\includegraphics[scale=0.4]{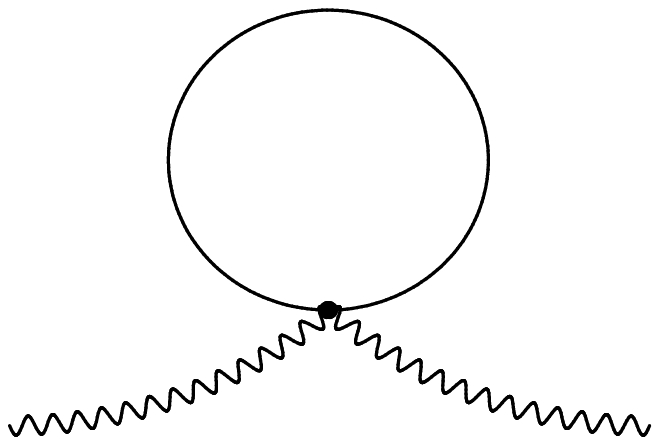}}
\end{picture}
\caption{The one-loop polarization operator of the quantum gauge superfield. The second column contains diagrams with a loop of the Faddeev--Popov ghosts, and the third one contains diagrams with a loop of the matter and Pauli--Villars superfields.}
\label{Figure_Polarization_Operator}
\end{figure}

\begin{figure}[h]
\hspace*{5mm}
\begin{picture}(0,5)
\put(0,4){$(1)$}
\put(0, 2.6){\includegraphics[scale=0.45]{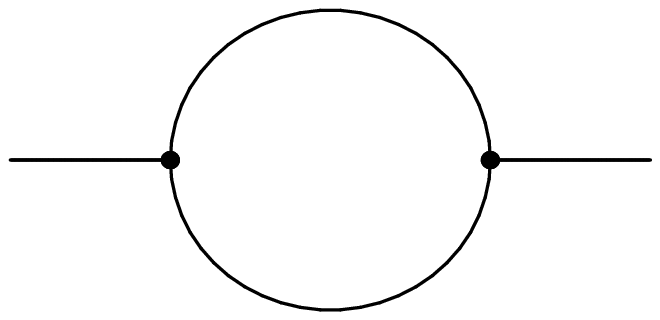}}
\put(4,4){$(2)$}
\put(4, 2.6){\includegraphics[scale=0.45]{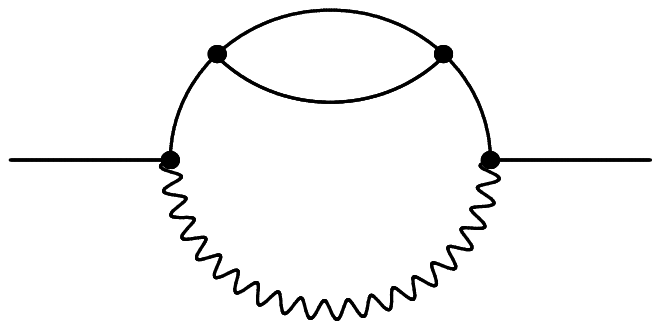}}
\put(8,4){$(3)$}
\put(8, 2.6){\includegraphics[scale=0.45]{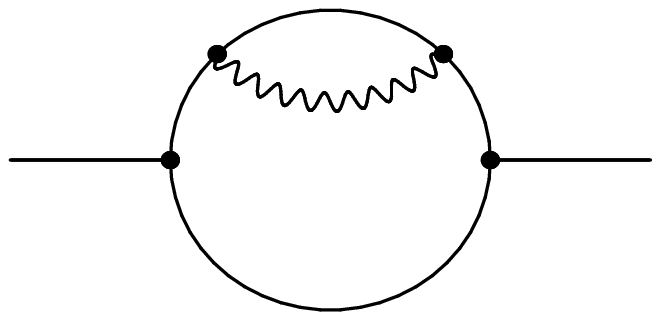}}
\put(12,4){$(4)$}
\put(12, 2.6){\includegraphics[scale=0.45]{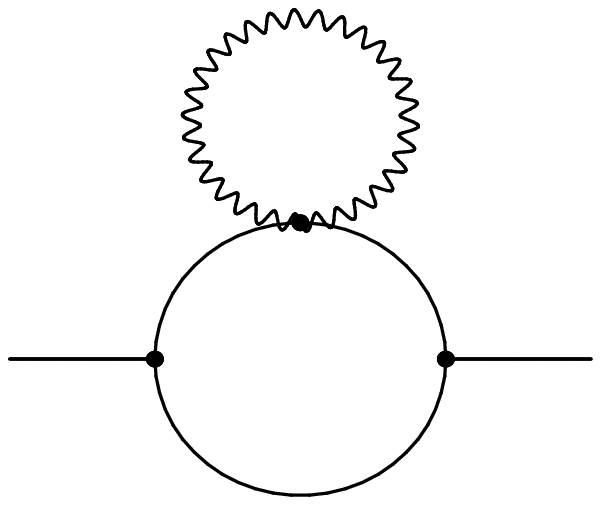}}
\put(0,2){$(5)$}
\put(0, 0.6){\includegraphics[scale=0.45]{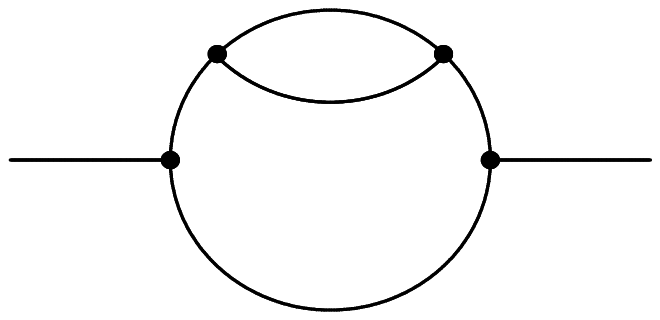}}
\put(4,2){$(6)$}
\put(4, 0.6){\includegraphics[scale=0.45]{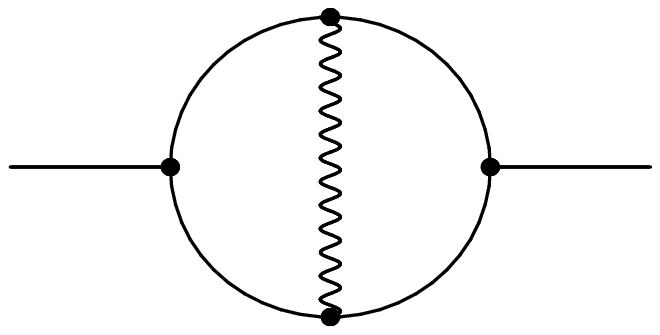}}
\put(8,2){$(7)$}
\put(8, 0.6){\includegraphics[scale=0.45]{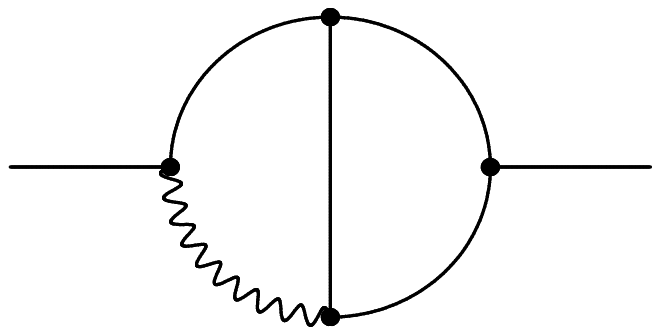}}
\put(12,2){$(8)$}
\put(12, 0.6){\includegraphics[scale=0.45]{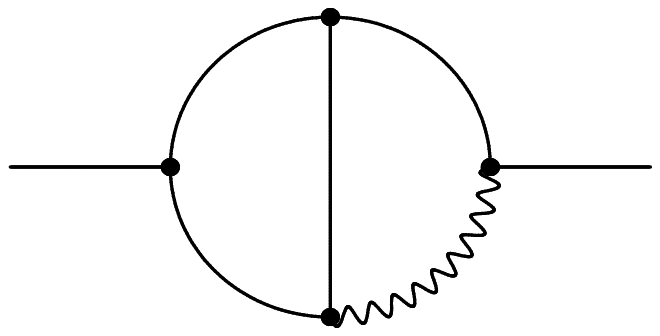}}
\end{picture}
\caption{Superdiagrams contributing to the two-loop anomalous dimension of the matter superfields which contain Yukawa vertices.}
\label{Figure_Yukawa_Diagrams}
\end{figure}

Superdiagrams contributing to the two-loop anomalous dimension of the matter superfields can be divided into two parts. The first part contains the superdiagrams without Yukawa vertices. They are presented in Fig. \ref{Figure_Diagrams_Without_Yukawa}. The gray circles in the superdiagrams (17) and (18) denote insertions of the one-loop polarization operator of the quantum gauge superfield, which is equal to the sum of the supergraphs presented in Fig. \ref{Figure_Polarization_Operator}. They have been calculated in Ref. \cite{Kazantsev:2017fdc}. The superdiagrams of the second part presented in Fig. \ref{Figure_Yukawa_Diagrams} contain the Yukawa vertices. They have already been calculated in Refs. \cite{Shakhmanov:2017soc,Kazantsev:2018nbl} for the higher derivative regulators $R(x)=1+x^m$ and $F(x)=1+x^n$. In this paper the integrals written in Refs. \cite{Shakhmanov:2017soc,Kazantsev:2018nbl} are calculated for an arbitrary form of the functions $R(x)$ and $F(x)$. The details of this calculation are presented in Appendix \ref{Appendix_Calculus}.

The superdiagrams presented in Fig. \ref{Figure_Diagrams_Without_Yukawa} have been calculated in Ref. \cite{Kataev:2017qvk} for ${\cal N}=1$ SQCD. It is essential that the action of ${\cal N}=1$ SQCD does not contain the Yukawa term cubic in the matter superfields, so that there is no need for the regulator $F(x)$. That is why the calculation of Ref.  \cite{Kataev:2017qvk} has been done for $F(x)=1$. However, for theories containing the Yukawa interaction the higher covariant derivative regularization should include $F(x)\ne 1$. In this paper we consider the general case. The presence of the function $F$ generates new vertices, which have to be taken into account. For example, the triple gauge-matter vertex is written as

\begin{equation}
\frac{1}{2}\int d^4x\, d^4\theta\, \phi^+ \bigg\{V F\Big(\frac{\partial^2}{\Lambda^2}\Big) + \sum\limits_{\alpha=1}^{\infty} f_\alpha \sum\limits_{\beta=0}^{\alpha-1}\Big(\frac{\partial^2}{\Lambda^2}\Big)^\beta\,\frac{\bar D^2[V, D^2]}{16\Lambda^2}\Big(\frac{\partial^2}{\Lambda^2}\Big)^{\alpha-1-\beta}\bigg\}\phi,
\end{equation}

\noindent
where the coefficients $f_\alpha$ are determined by the equation $F(x) = 1 +\sum\limits_{\alpha=1}^\infty f_\alpha x^\alpha$.

Remarkably, the sum of the supergraphs (1) --- (16) in Fig. \ref{Figure_Diagrams_Without_Yukawa} turns out to be independent of $F$. This regulator is present only in the superdiagrams (17) and (18) inside the expression for the polarization operator. Explicitly writing the sum of the superdiagrams depicted in Fig.~\ref{Figure_Diagrams_Without_Yukawa}, the two-loop anomalous dimension of the matter superfields can be presented as

\begin{eqnarray}\label{Gamma_Integral}
&& (\gamma_\phi)_i{}^j = \frac{d}{d\ln\Lambda} \bigg\{ -C(R)_i{}^j \int\frac{d^4k}{(2\pi)^4} \frac{2e_0^2}{k^4R_k} \Big[1 - \frac{2 e_0^2}{R_k}\Big(C_2 f(k/\Lambda)+T(R)h(k/\Lambda)\Big)\Big]\qquad\nonumber\\
&& + 4\left[C(R)^2\right]_i{}^j e_0^4 \int \frac{d^4k}{(2\pi)^4}\frac{d^4l}{(2\pi)^4} \bigg(\frac{1}{(l+k)^2R_{l+k}k^4R_kl^2} - \frac{1}{2l^4R_lk^4R_k} \bigg)\bigg\}_{\alpha,\lambda=\mbox{\scriptsize const}}\nonumber\\
&&\qquad\qquad\qquad\qquad\qquad\qquad\qquad\qquad\qquad  + \mbox{$\lambda$-dependent terms} + \mbox{higher orders}.\vphantom{\frac{1}{2}}
\end{eqnarray}

\noindent
The explicit expression for the $\lambda$-dependent terms can be found in Ref. \cite{Kazantsev:2018nbl}, see also Appendix~\ref{Appendix_Calculus}. The functions $f(k/\Lambda)$ and $h(k/\Lambda)$ are related to the one-loop polarization operator of the quantum gauge superfield in the Feynman gauge,

\begin{equation}\label{General_Result}
\Pi(\alpha_0,\lambda_0,k^2/\Lambda^2) = -8\pi\alpha_0 \Big(C_2\, f(k/\Lambda) + T(R)\,h(k/\Lambda)\Big) + O(\alpha_0^2,\alpha_0\lambda_0^2).
\end{equation}

\noindent
In our notation the polarization operator $\Pi(\alpha_0,\lambda_0,k^2/\Lambda^2)$ is defined by the equation

\begin{equation}\label{Polarization_Operator}
d_q^{-1}(\alpha_0,\lambda_0,k^2/\Lambda^2) - \alpha_0^{-1} R(k^2/\Lambda^2) \equiv - \alpha_0^{-1} \Pi(\alpha_0,\lambda_0,k^2/\Lambda^2),
\end{equation}

\noindent
where

\begin{equation}\label{Dq_Definition}
\Gamma^{(2)}_V - S^{(2)}_{\mbox{\scriptsize gf}}\equiv - \frac{1}{8\pi} \mbox{tr} \int \frac{d^4k}{(2\pi)^4}\, d^4\theta\, V(-k,\theta) \partial^2 \Pi_{1/2} V(k,\theta)\, d_q^{-1}\big(\alpha_0,\lambda_0,k^2/\Lambda^2\big).
\end{equation}

\noindent
The explicit expressions for the functions $f(k/\Lambda)$ and $h(k/\Lambda)$ are rather large. They can be found in Ref. \cite{Kazantsev:2017fdc}. However, in this paper we need only their asymptotic behavior at small $k$,

\begin{eqnarray}\label{F_Asymptotics}
&& f(k/\Lambda) = -\frac{3}{16\pi^2} \Big(\ln\frac{\Lambda}{k}+ \ln a_{\varphi} + 1 + o(1)\Big);\\
\label{H_Asymptotics}
&& h(k/\Lambda) = \frac{1}{16\pi^2} \Big(\ln\frac{\Lambda}{k} + \ln a + 1 + o(1)\Big),
\end{eqnarray}

\noindent where $o(1)$ denotes terms that vanish in the limit $k\to 0$.

In Eq. (\ref{Gamma_Integral}) the differentiation with respect to $\ln\Lambda$ is to be performed at fixed values of the renormalized couplings $\alpha$ and $\lambda$ before the momentum integration. This makes the integrals well-defined and finite in the infrared region.

The integrals giving the two-loop anomalous dimension are calculated in Appendix \ref{Appendix_Calculus}. We calculate both the integrals explicitly written in Eq. (\ref{Gamma_Integral}) and the integrals present in the $\lambda$-dependent terms for an arbitrary form of the functions $R(x)$ and $F(x)$. The result for the two-loop anomalous dimension defined in terms of the bare couplings obtained in Appendix \ref{Appendix_Calculus} is written as

\begin{eqnarray}\label{Gamma_Bare}
&&\hspace*{-7mm} (\gamma_\phi)_i{}^j(\alpha_0,\lambda_0) = - \frac{\alpha_0}{\pi}C(R)_i{}^j+\frac{1}{4\pi^2}\lambda^*_{0imn}\lambda_0^{jmn}  + \frac{\alpha_0^2}{2\pi^2} \left[C(R)^2\right]_i{}^j -\frac{1}{16\pi^4}\lambda^*_{0iac}\lambda_0^{jab}\lambda^*_{0bde}\lambda_0^{cde}\nonumber\\
&&\hspace*{-7mm} -\frac{3\alpha_0^2}{2\pi^2}\, C_2 C(R)_i{}^j\Big(\ln a_{\varphi}+1+\frac{A}{2}\Big) +\frac{\alpha_0^2}{2\pi^2}\, T(R) C(R)_i{}^j\Big(\ln a + 1 + \frac{A}{2}\Big) - \frac{\alpha_0}{8\pi^3} \lambda^*_{0lmn}\lambda^{jmn}_0 C(R)_i{}^l\nonumber\\
&&\hspace*{-7mm} \times (1-B+A) + \frac{\alpha_0}{4\pi^3}\lambda^*_{0imn}\lambda_0^{jml}C(R)_l{}^n(1-A+B)
+ O\Big(\alpha_0^3,\alpha_0^2\lambda_0^2,\alpha_0\lambda_0^4,\lambda_0^6\Big),
\end{eqnarray}

\noindent
where the constants $A$ and $B$ are given by the integrals

\begin{equation}\label{AB}
A=\int\limits_0^\infty dx \ln x\, \frac{d}{dx}\frac{1}{R(x)};\qquad B = \int\limits_0^\infty dx \ln x\, \frac{d}{dx}\frac{1}{F^2(x)}.
\end{equation}

\noindent
For the regulators $R(x)=1+x^m$ and $F(x)=1+x^n$ (which were used in Refs. \cite{Kazantsev:2018nbl,Soloshenko:2003sx}) these integrals can be taken,

\begin{equation}
A = 0;\qquad B =\frac{1}{n}.
\end{equation}

For the particular case of ${\cal N}=1$ SQED with $N_f$ flavors Eq. (\ref{Gamma_Bare}) gives

\begin{equation}\label{Gamma_Bare_SQED}
\gamma(\alpha_0) = -\frac{\alpha_0}{\pi} + \frac{\alpha_0^2}{\pi^2} \Big(N_f \ln a + N_f + \frac{N_f A}{2} + \frac{1}{2}\Big) + O(\alpha_0^3).
\end{equation}

\noindent
This expression agrees with the results of Refs. \cite{Kataev:2013csa,Soloshenko:2003sx,Aleshin:2020gec} obtained for $R(x)=1+x^m$ (and, therefore, $A=0$) and generalizes them to the case of an arbitrary regulator function. Note that according to Ref. \cite{Aleshin:2020gec} Eq. (\ref{Gamma_Bare_SQED}) is valid for an arbitrary $\xi$-gauge.

The anomalous dimension (\ref{Gamma_Bare}) is considerably simplified for theories finite in the one-loop approximation, which are obtained if

\begin{equation}\label{Finiteness_Conditions_Bare}
T(R) = 3C_2; \qquad \lambda^*_{0imn} \lambda_0^{jmn} = 4\pi\alpha_0 C(R)_i{}^j.
\end{equation}

\noindent
The first equation (which is an analog of the Banks--Zaks prescription \cite{Banks:1981nn}) leads to the vanishing one-loop $\beta$-function, while the second one follows from the vanishing of the one-loop anomalous dimension of the matter superfields. In this case we obtain

\begin{eqnarray}\label{Gamma_Finite_Bare}
&&\hspace*{-9mm} (\gamma_\phi)_i{}^j(\alpha_0,\lambda_0) = -\frac{3\alpha_0^2}{2\pi^2}C_2 C(R)_i{}^j \ln \frac{a_{\varphi}}{a} - \frac{\alpha_0}{4\pi^2}\Big(\frac{1}{\pi}\lambda^*_{0imn}\lambda_0^{jml}C(R)_l{}^n + 2 \alpha_0 \left[C(R)^2\right]_i{}^j\Big) (A-B)\nonumber\\
&&\hspace*{-9mm}  + O\Big(\alpha_0^3,\alpha_0^2\lambda_0^2,\alpha_0\lambda_0^4,\lambda_0^6\Big).
\end{eqnarray}

\noindent
This implies that the one-loop finite theories remain finite in the two-loop approximation if the higher derivative regularization is chosen in such a way that $A=B$ and $a = a_\varphi$. The first condition is automatically satisfied if $R(x)=F^2(x)$. That is why a version of the higher covariant derivative regularization leading to the two-loop finiteness is obtained by imposing the conditions

\begin{equation}\label{RF2}
R(x) =F^2(x)\qquad \mbox{and}\qquad  a=a_\varphi.
\end{equation}

\noindent
Possibly, this could help to reveal if the finiteness of the considered class of supersymmetric theories takes place in all loops and what is its underlying reason.

Another interesting particular case corresponds to the theories which satisfy the $P=\frac{1}{3} Q$ condition. In terms of the bare couplings it is written as

\begin{equation}\label{JJN_Constraint_Bare}
\lambda^*_{0imn}\lambda_0^{jmn} - 4\pi\alpha_0 C(R)_i{}^j = \frac{2\pi\alpha_0}{3} Q \delta_i{}^j.
\end{equation}

\noindent
In this case the anomalous dimension (\ref{Gamma_Bare}) takes the form

\begin{eqnarray}\label{Gamma_JJN_Bare}
&&\hspace*{-7mm} (\gamma_\phi)_i{}^j(\alpha_0,\lambda_0) = \Big(\frac{\alpha_0}{6\pi} Q - \frac{\alpha_0^2}{36\pi^2} Q^2\Big)\delta_i^j - \frac{\alpha_0}{4\pi^2} \Big(\frac{1}{\pi}\lambda^*_{0imn} \lambda_0^{jml} C(R)_l{}^n+2\alpha_0 \left[C(R)^2\right]_i{}^j \Big) (A-B)\nonumber\\
&&\hspace*{-7mm} -\frac{\alpha_0^2}{12\pi^2} Q C(R)_i{}^j (A-B) - \frac{3\alpha_0^2}{2\pi^2} C_2 C(R)_i{}^j \Big(\ln a_\varphi + \frac{1}{2} + \frac{A}{2}\Big)+ \frac{\alpha_0^2}{2\pi^2} T(R) C(R)_i{}^j \Big(\ln a + \frac{1}{2} + \frac{A}{2}\Big)\nonumber\\
&&\hspace*{-7mm}
+ O\Big(\alpha_0^3,\alpha_0^2\lambda_0^2,\alpha_0\lambda_0^4,\lambda_0^6\Big).\vphantom{\frac{1}{2}}
\end{eqnarray}

\noindent
From this result we see that in the considered approximation the anomalous dimension satisfies the exact JJN equation (\ref{Gamma_Exact}) if the parameters of the higher covariant derivative regularization satisfy the conditions

\begin{equation}\label{Regularization_For_JJN}
A=B;\qquad a =a_\varphi = \exp\Big(-\frac{1}{2}(A+1)\Big).
\end{equation}

\noindent
In particular, the first equality is valid for $R(x) = F^2(x)$, while the second one can be obtained by a special choice of the Pauli--Villars masses. In this case

\begin{eqnarray}\label{Gamma_JJN_Bare}
&& (\gamma_\phi)_i{}^j(\alpha_0,\lambda_0) = \Big(\frac{\alpha_0}{6\pi} Q - \frac{\alpha_0^2}{36\pi^2} Q^2\Big)\delta_i^j + O\Big(\alpha_0^3,\alpha_0^2\lambda_0^2,\alpha_0\lambda_0^4,\lambda_0^6\Big)\nonumber\\
&&\qquad\qquad\qquad\qquad\qquad\qquad\qquad = \frac{\alpha_0 Q}{6\pi (1+\alpha_0 Q/6\pi)}\delta_i^j + O\Big(\alpha_0^3,\alpha_0^2\lambda_0^2,\alpha_0\lambda_0^4,\lambda_0^6\Big).\qquad
\end{eqnarray}

\section{Two-loop anomalous dimension defined in terms of the renormalized couplings}
\hspace*{\parindent}\label{Section_Two_Loop_Gamma_Renormalized}

To obtain the anomalous dimension (standardly) defined in terms of the renormalized couplings, it is necessary to rewrite the left hand side of the equation (\ref{Gamma_Definition_Bare}) in terms of the renormalized couplings with the help of Eqs. (\ref{Two_Loop_Alpha}) and (\ref{One_Loop_Lambda}) and integrate the result with respect to $\ln\Lambda$. Then we obtain the function $(\ln Z_\phi)_i{}^j(\alpha,\lambda,\ln\Lambda/\mu)$, which should be rewritten in terms of the bare couplings using Eqs. (\ref{Two_Loop_Alpha}) and (\ref{One_Loop_Lambda}). According to Eq. (\ref{Gamma_Definition_Renormalized}), the result should be differentiated with respect to $\ln \mu$. The final expression for the anomalous dimension is obtained after expressing the bare couplings in terms of the renormalized ones again using Eqs. (\ref{Two_Loop_Alpha}) and (\ref{One_Loop_Lambda}). As a correctness check, one can verify cancellation of all $\ln\Lambda/\mu$, which follows from the general theory of the renormalization group (see, e.g., \cite{Vladimirov:1979ib}). The result of the above described calculation is written as

\begin{eqnarray}\label{Gamma_General}
&&\hspace*{-7mm} (\widetilde{\gamma}_\phi)_i{}^j(\alpha,\lambda) = - \frac{\alpha}{\pi}C(R)_i{}^j+\frac{1}{4\pi^2}\lambda^*_{imn}\lambda^{jmn} +\frac{\alpha^2}{2\pi^2}\left[C(R)^2\right]_i{}^j -\frac{1}{16\pi^4}\lambda^*_{iac}\lambda^{jab}\lambda^*_{bde}\lambda^{cde} \nonumber\\
&&\hspace*{-7mm} -\frac{3\alpha^2}{2\pi^2}C_2 C(R)_i{}^j\Big(\ln a_{\varphi}+1+\frac{A}{2}-b_{11}+g_{11}\Big) + \frac{\alpha^2}{2\pi^2} T(R)\,C(R)_i{}^j\Big(\ln a+1+\frac{A}{2}-b_{12}+g_{11}\Big) \nonumber\\
&&\hspace*{-7mm} -\frac{\alpha}{8\pi^3}\lambda^*_{lmn}\lambda^{jmn}C(R)_i{}^l\Big(1-B+A-2g_{12}+2g_{11}\Big) + \frac{\alpha}{4\pi^3}\lambda^*_{imn}\lambda^{jml}C(R)_l{}^n \Big(1-A+B +2g_{12}\nonumber\\
&&\hspace*{-7mm} -2g_{11}\Big) + O\Big(\alpha^3,\alpha^2\lambda^2,\alpha\lambda^4,\lambda^6\Big).
\end{eqnarray}

\noindent
It depends on the finite constants $b_i$ and $g_i$, which determine the renormalization prescription. Certainly, this agrees with the statement that the anomalous dimension defined in terms of the renormalized couplings is scheme dependent starting from the two-loop order. However,  the two-loop terms proportional to $\left[C(R)^2\right]_i{}^j$ and to $\lambda^*_{iac}\lambda^{jab}\lambda^*_{bde}\lambda^{cde}$
are scheme independent in agreement with Refs. \cite{Kataev:2014gxa} and \cite{Shakhmanov:2017soc}, respectively.

Now, let us consider particular cases of the general expression (\ref{Gamma_General}).

First, we write the expression for the anomalous dimension in the HD+MSL scheme, in which all finite constants (namely, $g_{11}$, $g_{12}$, $b_{11}$, and $b_{12}$ in the considered approximation) should be set to 0, so that

\begin{eqnarray}\label{Gamma_HD+MSL}
&&\hspace*{-7mm} (\widetilde{\gamma}_\phi)_i{}^j(\alpha,\lambda)_{\mbox{\scriptsize HD+MSL}} = - \frac{\alpha}{\pi}C(R)_i{}^j+\frac{1}{4\pi^2}\lambda^*_{imn}\lambda^{jmn} +\frac{\alpha^2}{2\pi^2}\left[C(R)^2\right]_i{}^j -\frac{1}{16\pi^4}\lambda^*_{iac}\lambda^{jab}\lambda^*_{bde}\lambda^{cde} \nonumber\\
&&\hspace*{-7mm} -\frac{3\alpha^2}{2\pi^2}C_2 C(R)_i{}^j\Big(\ln a_{\varphi}+1+\frac{A}{2}\Big) + \frac{\alpha^2}{2\pi^2} T(R)\,C(R)_i{}^j\Big(\ln a+1+\frac{A}{2}\Big) -\frac{\alpha}{8\pi^3}\lambda^*_{lmn}\lambda^{jmn}C(R)_i{}^l \nonumber\\
&&\hspace*{-7mm} \times \Big(1-B+A \Big) + \frac{\alpha}{4\pi^3}\lambda^*_{imn}\lambda^{jml}C(R)_l{}^n \Big(1-A+B\Big) + O\Big(\alpha^3,\alpha^2\lambda^2,\alpha\lambda^4,\lambda^6\Big).
\end{eqnarray}

\noindent
In agreement with the general statement \cite{Kataev:2013eta}, this expression coincides with Eq. (\ref{Gamma_Bare}) up to the renaming of arguments $\alpha\to\alpha_0$, $\lambda\to\lambda_0$.

Eq. (\ref{Gamma_General}) should also agree with the result in the $\overline{\mbox{DR}}$-scheme, which can be found, e.g., in Ref. \cite{Jack:1996vg}. This means that there must be such finite constants that the anomalous dimension takes the form

\begin{eqnarray}\label{Gamma_DR}
&&\hspace*{-7mm} (\widetilde{\gamma}_\phi)_i{}^j(\alpha,\lambda)_{\overline{\mbox{\scriptsize DR}}} = - \frac{\alpha}{\pi}C(R)_i{}^j+\frac{1}{4\pi^2}\lambda^*_{imn}\lambda^{jmn} +\frac{\alpha^2}{2\pi^2}\left[C(R)^2\right]_i{}^j -\frac{1}{16\pi^4}\lambda^*_{iac}\lambda^{jab}\lambda^*_{bde}\lambda^{cde} \nonumber\\
&&\hspace*{-7mm} -\frac{\alpha^2}{4\pi^2}\Big(3C_2 -T(R)\Big) C(R)_i{}^j -\frac{\alpha}{8\pi^3}\lambda^*_{lmn}\lambda^{jmn}C(R)_i{}^l + \frac{\alpha}{4\pi^3}\lambda^*_{imn}\lambda^{jml}C(R)_l{}^n \nonumber\\
&&\hspace*{-7mm} + O\Big(\alpha^3,\alpha^2\lambda^2,\alpha\lambda^4,\lambda^6\Big).\vphantom{\frac{1}{2}}
\end{eqnarray}

\noindent
This is really true for the finite constants satisfying the equations

\begin{equation}\label{DR_Constants}
\qquad b_{11} - g_{11} = \ln a_\varphi + \frac{1}{2}(1+A);\qquad b_{12} -g_{11} = \ln a + \frac{1}{2}(1+A);\qquad g_{12} - g_{11} = \frac{1}{2} (A-B).\qquad
\end{equation}

\noindent
The existence of these values can be considered as a non-trivial check of the calculation correctness. Note that the constant $g_{11}$ remains unfixed due to the arbitrariness of choosing the renormalization point $\mu$. Its value can be found by comparing the renormalized one-loop two-point Green functions for the matter superfields calculated with the higher covariant derivative regularization and in the $\overline{\mbox{DR}}$ scheme, see Appendix \ref{Appendix_Calculus} and Ref. \cite{Aleshin:2019yqj} for details. This gives

\begin{equation}\label{G11_Value}
g_{11} =  - \frac{1}{2} - \frac{A}{2}.
\end{equation}

For ${\cal N}=1$ SQED with $N_f$ flavors from Eq. (\ref{Gamma_General}) we obtain the expression

\begin{equation}\label{Gamma_Renormalized_SQED}
\widetilde\gamma(\alpha) = -\frac{\alpha}{\pi} + \frac{\alpha^2}{\pi^2} \Big(N_f \ln a + N_f + \frac{N_f A}{2} + \frac{1}{2} - N_f b_{12} + N_f g_{11}\Big) + O(\alpha^3),
\end{equation}

\noindent
which agrees with Ref. \cite{Kataev:2013csa}.

The ${\cal N}=1$ supersymmetric theories finite in the one-loop approximation are of a special interest. Written in terms of the renormalized couplings the finiteness conditions are

\begin{equation}\label{Finiteness_Conditions}
T(R) = 3C_2; \qquad \lambda^*_{imn} \lambda^{jmn} = 4\pi\alpha C(R)_i{}^j.
\end{equation}

\noindent
In this case the expression (\ref{Gamma_General}) takes the form

\begin{eqnarray}\label{Gamma_Finite_General}
&&\hspace*{-7mm} (\widetilde{\gamma}_\phi)_i{}^j(\alpha,\lambda) = -\frac{3\alpha^2}{2\pi^2}C_2 C(R)_i{}^j\Big(\ln \frac{a_{\varphi}}{a} -b_{11}+b_{12}\Big)
- \frac{\alpha}{4\pi^2} \Big(\frac{1}{\pi} \lambda^*_{imn} \lambda^{jml} C(R)_l{}^n + 2\alpha \left[C(R)^2\right]_i{}^j\Big) \nonumber\\
&&\hspace*{-7mm} \times \Big(A-B -2g_{12} +2g_{11}\Big) + O\Big(\alpha^3,\alpha^2\lambda^2,\alpha\lambda^4,\lambda^6\Big).
\end{eqnarray}

\noindent
We see that in general the one-loop finiteness does not lead to the two-loop finiteness. At first sight, this result contradicts the anomaly based consideration of Refs. \cite{Jones:1983vk,Jones:1984cx}. However, it seems that in Refs. \cite{Jones:1983vk,Jones:1984cx} the use of the $\overline{\mbox{DR}}$ scheme is implicitly assumed \cite{Jones:1986vp}. For this scheme the finite constants are given by Eq. (\ref{DR_Constants}), and the anomalous dimension (\ref{Gamma_Finite_General}) really vanishes. Also it is possible to find other subtraction schemes in which it is so. In particular, a more interesting example is the HD+MSL scheme obtained for the higher covariant derivative regularization (\ref{RF2}). The reason for this is that the HD+MSL scheme seems to be NSVZ in all loops. Moreover, the restrictions on the choice of the regularization can presumably reveal a deep structure of a theory needed for the finiteness, say, a certain symmetry underlying it.

One more interesting special case is the theories which satisfy the $P=\frac{1}{3}Q$ condition (\ref{JJN_Constraint}), under which the anomalous dimension is given by the expression

\begin{eqnarray}\label{Gamma_JJN}
&&\hspace*{-7mm} (\widetilde{\gamma}_\phi)_i{}^j(\alpha,\lambda) = \Big(\frac{\alpha}{6\pi} Q - \frac{\alpha^2}{36\pi^2} Q^2\Big)\delta_i^j + \Big(\frac{\alpha^2}{2\pi^2}\left[C(R)^2\right]_i{}^j
+ \frac{\alpha}{4\pi^3} \lambda^*_{imn} \lambda^{jml} C(R)_l{}^n\Big) \Big(-A+B\nonumber\\
&&\hspace*{-7mm} -2g_{11} + 2g_{12}\Big) -\frac{\alpha^2}{12\pi^2} Q C(R)_i{}^j \Big(A-B-2g_{12}+2g_{11}\Big) - \frac{3\alpha^2}{2\pi^2} C_2 C(R)_i{}^j \Big(\ln a_\varphi + \frac{1}{2}(1+A)\nonumber\\
&&\hspace*{-7mm} - b_{11} + g_{11}\Big)
+ \frac{\alpha^2}{2\pi^2} T(R) C(R)_i{}^j \Big(\ln a + \frac{1}{2}(1+A) - b_{12} + g_{11}\Big)
+ O\Big(\alpha^3,\alpha^2\lambda^2,\alpha\lambda^4,\lambda^6\Big).
\end{eqnarray}

\noindent
Again we see that there are two ways to obtain the JJN equation (\ref{Gamma_Exact}) in the considered approximation. Namely, it is possible to choose finite constants for which Eq. (\ref{DR_Constants}) is valid. The corresponding subtraction scheme is equivalent to the $\overline{\mbox{DR}}$ scheme. Another way is to use the HD+MSL scheme with the regularization satisfying Eq. (\ref{Regularization_For_JJN}).

\section{Three-loop $\beta$-function}
\label{Section_Three_Loop_Beta}

\subsection{The $\beta$-function defined in terms of the bare couplings}
\hspace*{\parindent}\label{Subsection_Three_Loop_Beta_Bare}

As we have already mentioned, there are strong indications that the NSVZ equation is valid in all loops for RGFs defined in terms of the bare couplings in the case of using the higher covariant derivative regularization. If we believe that this is really so, then it is possible to construct the expression for the three-loop $\beta$-function,

\begin{eqnarray}
&& \frac{\beta(\alpha_0,\lambda_0)}{\alpha_0^2} = - \frac{1}{2\pi} \Big(3C_2-T(R)\Big)\bigg(1+\frac{\alpha_0 C_2}{2\pi} + \frac{\alpha_0^2 C_2^2}{4\pi^2}\bigg) - \frac{1}{2\pi r} C(R)_j{}^i \big(\gamma_{\phi,\mbox{\scriptsize 1-loop}}\big)_i{}^j   \qquad\nonumber\\
&& - \frac{\alpha_0 C_2}{4\pi^2 r} C(R)_j{}^i \big(\gamma_{\phi,\mbox{\scriptsize 1-loop}}\big)_i{}^j - \frac{1}{2\pi r} C(R)_j{}^i \big(\gamma_{\phi,\mbox{\scriptsize 2-loop}}\big)_i{}^j + O\Big(\alpha_0^3,\alpha_0^2\lambda_0^2,\alpha_0\lambda_0^4,\lambda_0^6\Big),
\end{eqnarray}

\noindent
where $\big(\gamma_{\phi,\mbox{\scriptsize 1-loop}}\big)_i{}^j$ and $\big(\gamma_{\phi,\mbox{\scriptsize 2-loop}}\big)_i{}^j$ are the one- and two-loop parts of the anomalous dimension, respectively. Substituting these contributions we obtain

\begin{eqnarray}\label{Beta_Bare}
&&\hspace*{-7mm}  \frac{\beta(\alpha_0,\lambda_0)}{\alpha_0^2} = - \frac{1}{2\pi} \Big(3C_2-T(R)\Big) +\frac{\alpha_0}{4\pi^2} \Big\{ -3 C_2^2 + \frac{1}{r} C_2\, \mbox{tr}\,C(R) +\frac{2}{r}\, \mbox{tr}\left[C(R)^2\right] \Big\}\nonumber\\
&&\hspace*{-7mm}  - \frac{1}{8\pi^3 r} C(R)_j{}^i \lambda^*_{0imn} \lambda_0^{jmn}  + \frac{\alpha_0^2}{8\pi^3}\Big\{-3 C_2^3 + \frac{1}{r} C_2^2\, \mbox{tr}\, C(R)  - \frac{2}{r}\,\mbox{tr}\left[C(R)^3\right]  +\frac{2}{r} C_2\, \mbox{tr}\left[C(R)^2\right]\nonumber\\
&&\hspace*{-7mm} \times \Big(3\ln a_\varphi + 4 +\frac{3A}{2}\Big) - \frac{2}{r^2}\, \mbox{tr}\,C(R) \, \mbox{tr}\left[C(R)^2\right]\Big(\ln a + 1 +\frac{A}{2}\Big)\Big\} - \frac{\alpha_0 C_2}{16\pi^4 r} C(R)_j{}^i \lambda^*_{0imn} \lambda_0^{jmn}\nonumber\\
&&\hspace*{-7mm} + \frac{\alpha_0}{16\pi^4 r} \left[C(R)^2\right]_j{}^i \lambda^*_{0imn} \lambda_0^{jmn} \Big(1+A-B\Big) - \frac{\alpha_0}{8\pi^4 r} C(R)_j{}^i C(R)_l{}^n \lambda^*_{0imn} \lambda_0^{jml} \Big(1-A+B\Big) \nonumber\\
&&\hspace*{-7mm}  + \frac{1}{32\pi^5 r} C(R)_j{}^i \lambda^*_{0iac} \lambda_0^{jab} \lambda^*_{0bde} \lambda_0^{cde} + O\Big(\alpha_0^3,\,\alpha_0^2\lambda_0^2,\,\alpha_0\lambda_0^4,\,\lambda_0^6\Big). \vphantom{\frac{1}{\pi^2}}
\end{eqnarray}

\noindent
Certainly, this expression is independent of a renormalization prescription (for a fixed regularization) as any RGF defined in terms of the bare couplings. In the case of ${\cal N}=1$ SQED with $N_f$ flavors it gives

\begin{equation}\label{Beta_Bare_SQED}
\quad \frac{\beta(\alpha_0)}{\alpha_0^2} = \frac{N_f}{\pi} + \frac{\alpha_0 N_f}{\pi^2} + \frac{\alpha_0^2 N_f}{\pi^3}\Big[-\frac{1}{2} - N_f\Big(\ln a + 1 + \frac{A}{2}\Big)\Big] + O(\alpha_0^3) = \frac{N_f}{\pi}\Big(1-\gamma(\alpha_0)\Big) + O(\alpha_0^3).\quad
\end{equation}

For the one-loop finite theories (which satisfy Eq. (\ref{Finiteness_Conditions_Bare})) the expression (\ref{Beta_Bare}) is reduced to

\begin{eqnarray}\label{Beta_Bare_Finite}
&& \frac{\beta(\alpha_0,\lambda_0)}{\alpha_0^2} = \frac{\alpha_0}{8\pi^3 r} \Big(\frac{1}{\pi} C(R)_j{}^i C(R)_l{}^n \lambda^*_{0imn} \lambda_0^{jml} + 2\alpha_0\, \mbox{tr}\left[C(R)^3\right]\Big) (A-B)\qquad\nonumber\\
&&\qquad\qquad\qquad\qquad\qquad\qquad\quad + \frac{3\alpha_0^2}{4\pi^3 r} C_2\, \mbox{tr}\left[C(R)^2\right] \ln \frac{a_\varphi}{a} + O\Big(\alpha_0^3,\alpha_0^2\lambda_0^2,\alpha_0\lambda_0^4,\lambda_0^6\Big). \vphantom{\frac{1}{\pi^2}}\qquad
\end{eqnarray}

\noindent
We see that the three-loop $\beta$-function vanishes for the higher covariant derivative regularization with $A=B$ and $a_\varphi=a$, and, in particular, for the regularization (\ref{RF2}).

For theories satisfying the $P=\frac{1}{3}Q$ constraint (\ref{JJN_Constraint_Bare}) the three-loop $\beta$-function defined in terms of the bare couplings takes the form

\begin{eqnarray}\label{Beta_Bare_JJN}
&& \frac{\beta(\alpha_0,\lambda_0)}{\alpha_0^2} = \frac{1}{2\pi} Q - \frac{\alpha_0}{12\pi^2} Q^2 + \frac{\alpha_0^2}{72\pi^3} Q^3  + \frac{\alpha_0}{8\pi^3 r} \Big(\frac{1}{\pi} C(R)_j{}^i C(R)_l{}^n \lambda^*_{0imn} \lambda_0^{jml} \qquad \nonumber\\
&& +2\alpha_0\, \mbox{tr}\left[C(R)^3\right]\Big) (A-B) + \frac{3\alpha_0^2}{4\pi^3 r} C_2\, \mbox{tr}\left[C(R)^2\right] \ln \frac{a_\varphi}{a} + \frac{\alpha_0^2}{24\pi^3 r} Q\,\mbox{tr}\left[C(R)^2\right]  \nonumber\\
&& \times \Big(-6\ln a -3 -2A -B\Big) + O\Big(\alpha_0^3,\alpha_0^2\lambda_0^2,\alpha_0\lambda_0^4,\lambda_0^6\Big). \vphantom{\frac{\alpha_0^2}{\pi^2}}
\end{eqnarray}

\noindent
From this equation we see that in the considered approximation the JJN equation (\ref{Beta_JJN_Exact}) is satisfied by the $\beta$-function defined in terms of the bare couplings for the higher derivative regularization with the parameters obeying the conditions (\ref{Regularization_For_JJN}).

\subsection{The $\beta$-function defined in terms of the renormalized couplings}
\hspace*{\parindent}\label{Subsection_Three_Loop_Beta_Renormalized}

To find the $\beta$-function defined in terms of the renormalized couplings, first, it is necessary to rewrite the right hand side of Eq. (\ref{Beta_Bare}) in terms of the renormalized couplings and, then, integrate it
with respect to $\ln\Lambda$. The relations between the bare and renormalized gauge and Yukawa couplings are given by Eqs. (\ref{Two_Loop_Alpha}) and (\ref{One_Loop_Lambda}), respectively. After integrating Eq. (\ref{Beta_Bare}) with respect to $\ln\Lambda$ at fixed values of the renormalized couplings we find the three-loop expression for the renormalized gauge coupling constant as a function of the bare couplings. Next, the result is differentiated with respect to $\ln \mu$ (at fixed values of the bare couplings). Finally, the bare couplings are expressed in terms of the renormalized ones. Note that, according to the general theory of the renormalization group, all $\ln\Lambda/\mu$ should disappear. This fact can be used as a correctness check.

The procedure described above gives the three-loop $\beta$-function defined in terms of the renormalized couplings for a general renormalization prescription supplementing the higher covariant derivative regularization,

\begin{eqnarray}\label{Beta_General}
&&\hspace*{-7mm}  \frac{\widetilde\beta(\alpha,\lambda)}{\alpha^2} = - \frac{1}{2\pi} \Big(3C_2-T(R)\Big) +\frac{\alpha}{4\pi^2} \Big\{ -3 C_2^2 + \frac{1}{r} C_2\, \mbox{tr}\,C(R) +\frac{2}{r}\, \mbox{tr}\left[C(R)^2\right] \Big\} - \frac{1}{8\pi^3 r} C(R)_j{}^i \nonumber\\
&&\hspace*{-7mm}  \times\lambda^*_{imn} \lambda^{jmn}  + \frac{\alpha^2}{8\pi^3}\Big\{-3 C_2^3\Big(1+3b_{21}-3b_{11} \Big) + \frac{1}{r} C_2^2\, \mbox{tr}\, C(R)\Big(1+3b_{21}-3b_{12} +3b_{22} -3b_{11}\Big) \nonumber\\
&&\hspace*{-7mm}  - \frac{2}{r}\,\mbox{tr}\left[C(R)^3\right] - \frac{1}{r^2} C_2 \left[\,\mbox{tr}\, C(R)\right]^2 \Big(b_{22}-b_{12}\Big)+ \frac{1}{r} C_2\, \mbox{tr}\left[C(R)^2\right] \Big(6\ln a_\varphi +8 + 3A +6 b_{23} -6b_{11}\Big)\nonumber\\
&&\hspace*{-7mm}  + \frac{1}{r^2}\, \mbox{tr}\,C(R) \, \mbox{tr}\left[C(R)^2\right]\Big(-2\ln a -2 -A -2 b_{23} + 2 b_{12}\Big)\Big\} - \frac{\alpha C_2}{16\pi^4 r} C(R)_j{}^i \lambda^*_{imn} \lambda^{jmn}  + \frac{\alpha}{16\pi^4 r} \nonumber\\
&& \hspace*{-7mm} \times \left[C(R)^2\right]_j{}^i \lambda^*_{imn} \lambda^{jmn} \Big(1+A-B -2b_{24}+2g_{11}\Big) - \frac{\alpha}{8\pi^4 r} C(R)_j{}^i C(R)_l{}^n \lambda^*_{imn} \lambda^{jml} \Big(1-A+B\nonumber\\
&&\hspace*{-7mm} +2b_{24}-2g_{11}\Big) + \frac{1}{16\pi^5 r} C(R)_j{}^i \lambda^*_{iac} \lambda^{jab} \lambda^*_{bde} \lambda^{cde} \Big(\frac{1}{2} + b_{24} - g_{12}\Big) + \frac{1}{32\pi^5 r} C(R)_j{}^i \lambda^*_{imn} \lambda^{kmn} \lambda^*_{kpq}  \nonumber\\
&&\hspace*{-7mm} \times \lambda^{jpq} \Big(b_{24} - g_{12}\Big)  + O\Big(\alpha^3,\alpha^2\lambda^2,\alpha\lambda^4,\lambda^6\Big). \vphantom{\frac{1}{\pi^2}}
\end{eqnarray}

\noindent
We see that the terms corresponding to the three-loop contribution depend on the constants $b_i$ and $g_i$. Certainly, this follows from the fact that the $\beta$-function defined in terms of the renormalized couplings is scheme dependent starting from the three-loop approximation. The terms containing the Yukawa couplings exactly coincide with the ones obtained by the direct calculations of Refs. \cite{Shakhmanov:2017soc,Kazantsev:2018nbl}.

By definition, in the HD+MSL scheme all finite constants $b_i$ and $g_i$ are set to 0, so that the expression for the $\beta$-function takes the form

\begin{eqnarray}\label{Beta_HD+MSL}
&&\hspace*{-7mm}  \Big(\frac{\widetilde\beta(\alpha,\lambda)}{\alpha^2}\Big)_{\mbox{\scriptsize HD+MSL}} = - \frac{1}{2\pi} \Big(3C_2-T(R)\Big) +\frac{\alpha}{4\pi^2} \Big\{ -3 C_2^2 + \frac{1}{r} C_2\, \mbox{tr}\,C(R) +\frac{2}{r}\, \mbox{tr}\left[C(R)^2\right] \Big\}\nonumber\\
&&\hspace*{-7mm}  - \frac{1}{8\pi^3 r} C(R)_j{}^i \lambda^*_{imn} \lambda^{jmn}  + \frac{\alpha^2}{8\pi^3}\Big\{-3 C_2^3 + \frac{1}{r} C_2^2\, \mbox{tr}\, C(R)  - \frac{2}{r}\,\mbox{tr}\left[C(R)^3\right]  +\frac{1}{r} C_2\, \mbox{tr}\left[C(R)^2\right]\nonumber\\
&&\hspace*{-7mm} \times \Big(6\ln a_\varphi + 8 + 3A\Big) + \frac{1}{r^2}\, \mbox{tr}\,C(R) \, \mbox{tr}\left[C(R)^2\right]\Big(-2\ln a -2 -A\Big)\Big\} - \frac{\alpha C_2}{16\pi^4 r} C(R)_j{}^i \lambda^*_{imn} \lambda^{jmn}\nonumber\\
&&\hspace*{-7mm} + \frac{\alpha}{16\pi^4 r} \left[C(R)^2\right]_j{}^i \lambda^*_{imn} \lambda^{jmn} \Big(1+A-B\Big) - \frac{\alpha}{8\pi^4 r} C(R)_j{}^i C(R)_l{}^n \lambda^*_{imn} \lambda^{jml} \Big(1-A+B\Big) \nonumber\\
&&\hspace*{-7mm}  + \frac{1}{32\pi^5 r} C(R)_j{}^i \lambda^*_{iac} \lambda^{jab} \lambda^*_{bde} \lambda^{cde} + O\Big(\alpha^3,\,\alpha^2\lambda^2,\,\alpha\lambda^4,\,\lambda^6\Big) \vphantom{\frac{1}{\pi^2}}
\end{eqnarray}

\noindent
and up to the renaming of arguments $\alpha\to \alpha_0$, $\lambda\to \lambda_0$ coincides with the $\beta$-function defined in terms of the bare couplings given by Eq. (\ref{Beta_Bare}).

For certain values of the finite constants $b_i$ and $g_i$ the expression (\ref{Beta_General}) should also reproduce the $\overline{\mbox{DR}}$-result first found in Refs. \cite{Jack:1996vg,Jack:1996cn}, which in our notation takes the form

\begin{eqnarray}\label{Beta_DR}
&&\hspace*{-7mm}  \Big(\frac{\widetilde\beta(\alpha,\lambda)}{\alpha^2}\Big)_{\overline{\mbox{\scriptsize DR}}} = - \frac{1}{2\pi} \Big(3C_2-T(R)\Big) +\frac{\alpha}{4\pi^2} \Big\{ -3 C_2^2 + \frac{1}{r} C_2\, \mbox{tr}\, C(R) +\frac{2}{r}\, \mbox{tr}\left[C(R)^2\right] \Big\}\nonumber\\
&&\hspace*{-7mm}  - \frac{1}{8\pi^3 r} C(R)_j{}^i \lambda^*_{imn} \lambda^{jmn}  + \frac{\alpha^2}{8\pi^3}\Big\{-\frac{21}{4} C_2^3 + \frac{5}{2r} C_2^2\, \mbox{tr}\, C(R) - \frac{1}{4r^2} C_2 \left[\,\mbox{tr}\, C(R)\right]^2 - \frac{2}{r}\,\mbox{tr}\left[C(R)^3\right] \nonumber\\
&&\hspace*{-7mm}    + \frac{13}{2r} C_2\, \mbox{tr}\left[C(R)^2\right]   - \frac{3}{2r^2} \mbox{tr}\, C(R)\, \mbox{tr}\left[C(R)^2\right]\Big\} - \frac{\alpha C_2}{16\pi^4 r} C(R)_j{}^i \lambda^*_{imn} \lambda^{jmn}
+\frac{\alpha}{32\pi^4 r} \left[C(R)^2\right]_j{}^i \nonumber\\
&&\hspace*{-7mm}
\times \lambda^*_{imn} \lambda^{jmn} - \frac{3\alpha}{16\pi^4 r} C(R)_j{}^i  C(R)_l{}^n \lambda^*_{imn} \lambda^{jml} + \frac{3}{64\pi^4 r} C(R)_j{}^i \lambda^*_{iac} \lambda^{jab} \lambda^*_{bde} \lambda^{cde}
+ \frac{1}{128\pi^5 r} \nonumber\\
&&\hspace*{-7mm} \times C(R)_j{}^i \lambda^*_{imn} \lambda^{kmn} \lambda^*_{kpq} \lambda^{jpq} + O\Big(\alpha^3,\alpha^2\lambda^2,\alpha\lambda^4,\lambda^6\Big). \vphantom{\frac{1}{\pi^2}}
\end{eqnarray}

\noindent
In particular, this implies that all scheme independent terms in Eq. (\ref{Beta_General}) should coincide with the corresponding terms in Eq. (\ref{Beta_DR}). We see that it is really so. The other terms coincide if

\begin{eqnarray}
&& b_{21} - b_{11} = \frac{1}{4};\qquad b_{21} - b_{12} + b_{22} - b_{11} = \frac{1}{2};\qquad b_{22} - b_{12} = \frac{1}{4}; \nonumber\\
&& b_{23}-b_{12} = - \ln a - \frac{A}{2} -\frac{1}{4};\qquad\quad b_{23} - b_{11} = - \ln a_\varphi -\frac{A}{2} -\frac{1}{4};\nonumber\\
&& b_{24} -g_{11} = \frac{A}{2}-\frac{B}{2}+\frac{1}{4};\qquad\qquad\  b_{24} - g_{12} =\frac{1}{4}.
\end{eqnarray}

\noindent
From these equations and Eq. (\ref{DR_Constants}) one can express all finite constants in terms of $g_{11}$ given by Eq. (\ref{G11_Value}),

\begin{eqnarray}\label{DR_BG_Constants}
&& b_{11} = g_{11} + \ln a_\varphi + \frac{1}{2}(1+A) = \ln a_\varphi;\qquad\quad b_{12} = g_{11} + \ln a + \frac{1}{2}(1+A) = \ln a;\nonumber\\
&& b_{21} =  g_{11} + \ln a_\varphi + \frac{3}{4} +\frac{A}{2} = \ln a_\varphi + \frac{1}{4};\qquad\, b_{22} = g_{11} + \ln a + \frac{3}{4} +\frac{A}{2} = \ln a + \frac{1}{4};\qquad\nonumber\\
&& b_{23} = g_{11} +\frac{1}{4} = -\frac{1}{4}-\frac{A}{2};\qquad\qquad\qquad\qquad b_{24} = g_{11} + \frac{A}{2}-\frac{B}{2}+\frac{1}{4} = -\frac{1}{4}-\frac{B}{2};\qquad\nonumber\\
&& g_{12} = g_{11} + \frac{1}{2}(A-B) = -\frac{1}{2} -\frac{B}{2};\qquad\qquad\   g_{11} = -\frac{1}{2} - \frac{A}{2}.
\end{eqnarray}

\noindent
The values (\ref{DR_BG_Constants}) of the finite constants $b_i$ and $g_i$ correspond to the $\overline{\mbox{DR}}$ renormalization scheme.

Comparing Eqs. (\ref{Beta_General}) and (\ref{Gamma_General}) we see that for a general renormalization prescription the NSVZ relation does not take place. However, if the finite constants $b_i$ and $g_i$ satisfy the equations

\begin{eqnarray}\label{Finite_Constants_NSVZ}
&& b_{21} = b_{11};\qquad b_{22} = b_{12}; \qquad b_{23} = g_{11}; \qquad b_{24} = g_{12},
\end{eqnarray}

\noindent
then the NSVZ $\beta$-function in the considered approximation is valid for RGFs defined in terms of the renormalized couplings. Evidently, Eq. (\ref{Finite_Constants_NSVZ}) does not unambiguously fix the values of all finite constants. This implies that there is a class of NSVZ schemes similar to the one in the Abelian case which was described in Ref. \cite{Goriachuk:2018cac}.

From Eq. (\ref{Beta_General}) we see that the $\beta$-function of ${\cal N}=1$ SQED with $N_f$ flavors is given by the expression

\begin{equation}
\frac{\widetilde\beta(\alpha)}{\alpha^2} = \frac{N_f}{\pi} + \frac{\alpha N_f}{\pi^2} + \frac{\alpha^2 N_f}{\pi^3}\Big[-\frac{1}{2} + N_f\Big(-\ln a - 1 -\frac{A}{2} - b_{23} + b_{12}\Big)\Big]+ O(\alpha^3),
\end{equation}

\noindent
which agrees with Ref. \cite{Kataev:2013csa}.

For the one-loop finite theories (which satisfy the conditions (\ref{Finiteness_Conditions})) the three-loop $\beta$-function defined in terms of the renormalized couplings takes the form

\begin{eqnarray}\label{Beta_Renormalized_Finite}
&& \frac{\widetilde\beta(\alpha,\lambda)}{\alpha^2} = \frac{3\alpha^2}{4\pi^3 r} C_2\, \mbox{tr}\left[C(R)^2\right] \Big(\ln \frac{a_\varphi}{a} + b_{12} - b_{11}\Big) + \frac{\alpha}{8\pi^3 r} \Big(\frac{1}{\pi} C(R)_j{}^i C(R)_l{}^n \lambda^*_{imn} \lambda^{jml} \qquad\nonumber\\
&&  +2\alpha\,\mbox{tr}\left[C(R)^3\right]\Big) \Big(A-B -2g_{12}+2g_{11}\Big) + O\Big(\alpha^3,\alpha^2\lambda^2,\alpha\lambda^4,\lambda^6\Big). \vphantom{\frac{1}{\pi^2}}
\end{eqnarray}

\noindent
Comparing this expression with Eq. (\ref{Gamma_Finite_General}) we see that in this case the NSVZ equation

\begin{equation}
\frac{\widetilde\beta(\alpha,\lambda)}{\alpha^2} = - \frac{C(R)_j{}^i \big(\widetilde\gamma_\phi\big)_i{}^j(\alpha,\lambda)}{2\pi r\big(1 - \alpha C_2/2\pi\big)}
\end{equation}

\noindent
is valid in the considered approximation for a general renormalization prescription in agreement with the general statement that for a theory finite in a certain loop the $\beta$-function always vanishes in the next loop \cite{Grisaru:1985tc}. Note that in Ref. \cite{Grisaru:1985tc} the $\overline{\mbox{DR}}$ scheme was essentially used. However, in general, the NSVZ equation is not valid in the $\overline{\mbox{DR}}$ scheme (see, e.g., \cite{Jack:1996vg}). Now we see how this seeming contradiction is solved in the considered approximation.

Also we see that for theories satisfying Eq. (\ref{Finiteness_Conditions}) the three-loop $\beta$-function vanishes if

\begin{equation}
\ln\frac{a_\varphi}{a} + b_{12} - b_{11}  = 0;\qquad A-B-2g_{12} + 2g_{11} = 0.
\end{equation}

\noindent
In particular, these equations are valid in the $\overline{\mbox{DR}}$ scheme and in the HD+MSL scheme supplementing the regularization with $R(x)=F^2(x)$ and $a_\varphi=a$.

For theories which satisfy the $P=\frac{1}{3}Q$ condition (\ref{JJN_Constraint}) the $\beta$-function defined in terms of the renormalized couplings is

\begin{eqnarray}\label{Beta_Renormalized_JJN}
&&\hspace*{-7mm} \frac{\widetilde\beta(\alpha,\lambda)}{\alpha^2} = \frac{1}{2\pi} Q - \frac{\alpha}{12\pi^2} Q^2 + \frac{\alpha^2}{72\pi^3} Q^3 \Big(1+3b_{24}-3g_{12}\Big) + \frac{\alpha^2}{8\pi^3}C_2 Q^2\Big(b_{24}-b_{22}+b_{12}-g_{12}\Big) \nonumber\\
&&\hspace*{-7mm} + \frac{3\alpha^2}{8\pi^3} C_2^2 Q \Big(b_{21}-b_{11}-b_{22}+b_{12}\Big) + \frac{\alpha}{8\pi^3 r} \Big(\frac{1}{\pi} C(R)_j{}^i C(R)_l{}^n \lambda^*_{imn} \lambda^{jml} + 2\alpha\, \mbox{tr}\left[C(R)^3\right] \Big)\nonumber\\
&&\hspace*{-7mm} \times \Big(A-B - 2g_{12}+2g_{11}\Big)+ \frac{\alpha^2}{24\pi^3 r} Q\,\mbox{tr}\left[C(R)^2\right] \Big(-6\ln a - 3 -2A -B - 6b_{23} + 6b_{12} + 6 b_{24}\nonumber\\
&&\hspace*{-7mm}  +2g_{11} -8 g_{12}\Big) + \frac{3\alpha^2}{4\pi^3 r} C_2 \mbox{tr}\left[C(R)^2\right]\Big(\ln \frac{a_\varphi}{a} +b_{12} -b_{11}\Big)
+ O\Big(\alpha^3,\alpha^2\lambda^2,\alpha\lambda^4,\lambda^6\Big). \vphantom{\frac{1}{\pi^2}}
\end{eqnarray}

\noindent
Therefore, Eq. (\ref{Beta_JJN_Exact}) in the considered approximation is valid, for example, in the HD+MSL scheme with the regularization satisfying the conditions (\ref{Regularization_For_JJN}). Another possibility is the NSVZ scheme constructed from the $\overline{\mbox{DR}}$ scheme by a special redefinition of the gauge coupling constant in Refs. \cite{Jack:1996vg,Jack:1996cn,Jack:1998uj}. This scheme corresponds to the finite constants

\begin{eqnarray}\label{DR_NSVZ_BG_Constants}
&& b_{21} = b_{11} = \ln a_\varphi;\qquad\qquad\quad b_{22} = b_{12} = \ln a;\qquad \nonumber\\
&& b_{23} = g_{11} = -\frac{1}{2} - \frac{A}{2};\quad\qquad\ b_{24} = g_{12} = -\frac{1}{2} - \frac{B}{2}.\qquad
\end{eqnarray}

\section{Conclusion}
\hspace*{\parindent}

In this paper the two-loop anomalous dimension of the matter superfields has been calculated for a general ${\cal N}=1$ supersymmetric gauge theory regularized by higher covariant derivatives. The result has been found both for the anomalous dimension defined in terms of the bare couplings and for the one defined in terms of the renormalized couplings for an arbitrary renormalization prescription. For theories finite in the one-loop approximation the two-loop anomalous dimension defined in terms of the bare couplings does not in general vanish. However, for a version of the higher derivative regularization with the parameters $R(x) = F^2(x)$ and $a=a_\varphi$ this is so. Possibly, this could help to understand deeper reasons responsible for the finiteness of a theory. Moreover, we analysed a possibility of satisfying the exact equation (\ref{Gamma_Exact}) for the anomalous dimension which was proposed by Jack, Jones, and North for theories obeying the $P=\frac{1}{3}Q$ condition. It appears that in the considered approximation for this equation to be valid the regularization parameters should satisfy the constraints (\ref{Regularization_For_JJN}).

Using the statement that for RGFs defined in terms of the bare couplings the NSVZ equation is valid in the case of using the higher covariant derivative regularization we also construct the expression for the three-loop $\beta$-function. Again, we present the results for the $\beta$-function defined in terms of the bare couplings and for the one defined in terms of the renormalized couplings. For the one-loop finite theories the three-loop $\beta$-function defined in terms of the bare couplings vanishes under the same conditions (\ref{RF2}) as the two-loop anomalous dimension. For theories satisfying Eq. (\ref{JJN_Constraint}) the exact expression for the $\beta$-function in the form of the geometric series (\ref{Beta_JJN_Exact}) is valid with the regularization (\ref{Regularization_For_JJN}).

For RGFs defined in terms of the renormalized couplings the above results are valid in the HD+MSL scheme. Also we have demonstrated that for the one-loop finite theories the NSVZ equation in the considered approximation is valid for an arbitrary renormalization prescription.

RGFs obtained in this paper with the help of the higher covariant derivative regularization are in agreement with the ones in the $\overline{\mbox{DR}}$ scheme in a sense that there are certain values of finite constants fixing the renormalization prescription for which the results of this paper reproduce the $\overline{\mbox{DR}}$ results.

\section*{Acknowledgments}
\hspace*{\parindent}

The authors are very grateful to A.L.Kataev for valuable discussions. Also the authors would like to thank A.B.Pimenov and M.B.Skoptsov for participating in the initial stage of this work and A.A.Soloshenko for discussions that started it.

The work was supported by the Foundation for the Advancement of Theoretical Physics and Mathematics ``BASIS'', grants no 17-11-120 (A.K.) and 19-1-1-45-1 (K.S.).

\appendix

\section*{Appendix}

\section{Calculation of the loop integrals giving the two-loop anomalous dimension}
\hspace*{\parindent}\label{Appendix_Calculus}

The two-loop anomalous dimension can be presented in the form

\begin{eqnarray}\label{Two_Loop_Gamma_Integrals}
&&\hspace*{-6mm} (\gamma_\phi)_i{}^j(\alpha_0,\lambda_0)= - 8 \pi \alpha C(R)_i{}^j I_1 + 2 \lambda^*_{imn} \lambda^{jmn} I_2 - 64 \pi^2 \alpha^2 \left[C(R)^2\right]_i{}^j I_{10}
+ 64 \pi^2 \alpha^2 C_2 C(R)_i{}^j \vphantom{\frac{1}{2}}\nonumber\\
&&\hspace*{-6mm} \times \bigg[I_{11} + \frac{3}{16\pi^2}\Big(\ln\frac{\Lambda}{\mu}+b_{11}\Big) I_1 - \frac{3}{2} I_8\bigg]
+ 64 \pi^2 \alpha^2 C(R)_i{}^j T(R) \bigg[I_{12} - \frac{1}{16\pi^2} \Big(\ln\frac{\Lambda}{\mu} + b_{12}\Big) I_1 \nonumber\\
&&\hspace*{-6mm} + \frac{1}{2} I_8 \bigg] +16 \pi \alpha \lambda^*_{lmn} \lambda^{jmn} C(R)_i{}^l \bigg[I_7 - I_4 -I_8+\frac{1}{8\pi^2}\Bigl(\ln\frac{\Lambda}{\mu}+g_{11}\Bigr)I_1\bigg]
+ 32 \pi \alpha \lambda^*_{imn} \lambda^{jml} C(R)_l{}^n \nonumber\\
&&\hspace*{-6mm} \times \bigg[I_3+I_9 -\frac{1}{8\pi^2}\Big(\ln\frac{\Lambda}{\mu}+g_{11}\Big) I_2\bigg]
- 2 \lambda^*_{iab} \lambda^{kab} \lambda^*_{kcd} \lambda^{jcd} I_6 - 8 \lambda^*_{iac} \lambda^{jab} \lambda^*_{bde} \lambda^{cde} \bigg[I_5+I_9-\frac{1}{8\pi^2}\nonumber\\
&&\hspace*{-6mm} \times \Big(\ln\frac{\Lambda}{\mu}+g_{12}\Big)I_2\bigg] + O\Big(\alpha^3,\alpha^2\lambda^2,\alpha\lambda^4,\lambda^6\Big),
\end{eqnarray}

\noindent where the Euclidean integrals $I_1$ --- $I_{12}$ are given by the expressions

\begin{eqnarray}\label{I1_Definition}
&&\hspace*{-3mm} I_1 \equiv \frac{d}{d\ln\Lambda}\int\frac{d^4k}{(2\pi)^4} \frac{1}{k^4R_k} = \frac{1}{8\pi^2};\\
&&\hspace*{-3mm} I_2 \equiv \frac{d}{d\ln\Lambda}\int\frac{d^4k}{(2\pi)^4} \frac{1}{k^4F_k^2} = \frac{1}{8\pi^2};\\
&&\hspace*{-3mm} I_3 \equiv \frac{d}{d\ln\Lambda}\int\frac{d^4l}{(2\pi)^4} \frac{1}{l^4F_l^2}\bigg(\int\frac{d^4k}{(2\pi)^4} \frac{1}{k^2R_k(l+k)^2}-\frac{1}{8\pi^2}\ln\frac{\Lambda}{l}\bigg);\\
&&\hspace*{-3mm} I_4 \equiv \frac{d}{d\ln\Lambda}\int\frac{d^4k}{(2\pi)^4} \frac{1}{k^4R_k}\bigg(\int\frac{d^4l}{(2\pi)^4} \frac{1}{l^2F_l(l+k)^2F_{l+k}}-\frac{1}{8\pi^2}\ln\frac{\Lambda}{k}\bigg);\\
&&\hspace*{-3mm} I_5 \equiv \frac{d}{d\ln\Lambda}\int\frac{d^4k}{(2\pi)^4} \frac{1}{k^4F_k^3}\bigg(\int\frac{d^4l}{(2\pi)^4} \frac{1}{l^2F_l(l+k)^2F_{l+k}}-\frac{1}{8\pi^2}F_k\ln\frac{\Lambda}{k}\bigg);\\
&&\hspace*{-3mm} I_6 \equiv \frac{d}{d\ln\Lambda}\bigg[\int\frac{d^4k}{(2\pi)^4} \frac{d^4l}{(2\pi^4)} \frac{1}{k^4F_k^2l^4F_l^2}
-\frac{1}{4\pi^2}\Big(\ln\frac{\Lambda}{\mu}+g_{12}\Big)\int\frac{d^4k}{(2\pi)^4} \frac{1}{k^4F_k^2}\bigg];\\
&&\hspace*{-3mm} I_7 \equiv \frac{d}{d\ln\Lambda}\bigg[\int\frac{d^4k}{(2\pi)^4} \frac{d^4l}{(2\pi)^4}\frac{1}{k^4R_kl^4F_l^2}
-\frac{1}{8\pi^2}\Big(\ln\frac{\Lambda}{\mu}+g_{11}\Big)\int\frac{d^4k}{(2\pi)^4} \Big(\frac{1}{k^4F_k^2} + \frac{1}{k^4R_k}\Big)\bigg];\qquad\\
&&\hspace*{-3mm} I_8 \equiv \frac{1}{8\pi^2}\int\frac{d^4k}{(2\pi)^4}\frac{1}{k^4}\, \ln\frac{\Lambda}{k}\frac{d}{d\ln\Lambda}\frac{1}{R_k};\\
&&\hspace*{-3mm} I_9 \equiv \frac{1}{8\pi^2}\int\frac{d^4k}{(2\pi)^4}\frac{1}{k^4}\, \ln\frac{\Lambda}{k}\frac{d}{d\ln\Lambda}\frac{1}{F_k^2};\\
&&\hspace*{-3mm} I_{10} \equiv \frac{d}{d\ln\Lambda}\int\frac{d^4k}{(2\pi)^4} \frac{d^4l}{(2\pi)^4}\frac{k_\mu l^\mu}{l^4R_lk^4R_k(l+k)^2};\\
&&\hspace*{-3mm} I_{11} \equiv \frac{d}{d\ln\Lambda}\int\frac{d^4k}{(2\pi)^4} \frac{1}{k^4R_k^2}\bigg[f(k/\Lambda) + \frac{3R_k}{16\pi^2}\ln\frac{\Lambda}{k}\bigg];\\
\label{I12_Definition}
&&\hspace*{-3mm} I_{12} \equiv \frac{d}{d\ln\Lambda}\int\frac{d^4k}{(2\pi)^4} \frac{1}{k^4R_k^2}\bigg[h(k/\Lambda) - \frac{R_k}{16\pi^2}\ln\frac{\Lambda}{k}\bigg].
\end{eqnarray}

\noindent
The (rather large) explicit expressions for the functions $f(k/\Lambda)$ and $h(k/\Lambda)$ inside the integrals $I_{11}$ and $I_{12}$ can be found in Ref. \cite{Kazantsev:2017fdc}.

The integrals $I_6$, $I_7$, and $I_{10}$ have been calculated in Refs. \cite{Shakhmanov:2017soc}, \cite{Kazantsev:2018nbl}, and \cite{Kataev:2019olb}, respectively. The results are written as

\begin{equation}
I_6 = -\frac{1}{32\pi^4}\Big(\ln\frac{\Lambda}{\mu} + g_{12}\Big);\qquad I_7 = -\frac{1}{32\pi^4}\Big(\ln\frac{\Lambda}{\mu} + g_{11}\Big); \qquad I_{10} = -\frac{1}{128\pi^4}.
\end{equation}

\noindent
Calculating the integrals $I_8$ and $I_9$ in the four-dimensional spherical coordinates one can relate them to the constants $A$ and $B$ defined by Eq. (\ref{AB}),

\begin{equation}
I_8 = \frac{A}{128\pi^4};\qquad I_9 = \frac{B}{128\pi^4}.
\end{equation}

Earlier the integrals $I_3$, $I_4$, $I_5$ and the integrals $I_{11}$, $I_{12}$ have been found only for the higher derivative regulators $R(x)=1+x^m$ and  $F(x)=1+x^n$, see Refs. \cite{Kazantsev:2018nbl} and \cite{Kazantsev:2018kjx}, respectively. However, it is possible to generalize the results to the case of arbitrary functions $R(x)$ and $F(x)$. For this purpose we will use the equation

\begin{equation}\label{Auxiliary_A_Equation}
\frac{d}{d\ln\Lambda} \int \frac{d^4k}{(2\pi)^4} \frac{a(k/\Lambda)}{k^4} = \frac{1}{8\pi^2} a(0)
\end{equation}

\noindent
valid for a nonsingular function $a(k/\Lambda)$ rapidly decreasing at infinity. Using this equation the integrals under consideration can be presented in the form

\begin{eqnarray}
\label{I3}
&& I_3 = \lim\limits_{l\to 0}\, \frac{1}{8\pi^2} \Big( \int\frac{d^4k}{(2\pi)^4} \frac{1}{k^2R_k(l+k)^2} - \frac{1}{8\pi^2} \ln\frac{\Lambda}{l} \Big);\\
\label{I4_I5}
&& I_4 = I_5 = \lim\limits_{k\to 0}\, \frac{1}{8\pi^2} \Big( \int\frac{d^4l}{(2\pi)^4} \frac{1}{l^2F_l(l+k)^2F_{l+k}} - \frac{1}{8\pi^2} \ln\frac{\Lambda}{k} \Big);\\
\label{I11}
&& I_{11} = \lim\limits_{k\to 0}\, \frac{1}{8\pi^2}\Big(f(k/\Lambda)+\frac{3}{16\pi^2}\ln\frac{\Lambda}{k}\Big) = -\frac{3}{128\pi^4}\Big(\ln a_\varphi+1\Big);\\
\label{I12}
&& I_{12} = \lim\limits_{k\to 0}\, \frac{1}{8\pi^2}\Big(h(k/\Lambda)-\frac{1}{16\pi^2}\ln\frac{\Lambda}{k}\Big) = \frac{1}{128\pi^4}\Big(\ln a+1\Big),
\end{eqnarray}

\noindent
where the last equalities in Eqs. (\ref{I11}) and (\ref{I12}) follow from Eqs. (\ref{F_Asymptotics}) and (\ref{H_Asymptotics}), respectively.

Now, let us calculate the integral $I_3$ for a general higher derivative regulator $R(x)$. Introducing the Feynman parameter $x$ and making the substitution $k_\mu \to k_\mu - x l_\mu$ the integral in Eq. (\ref{I3}) can be rewritten as

\begin{eqnarray}\label{Integral_For_I3}
&& \int\frac{d^4k}{(2\pi)^4}\frac{1}{k^2R_k(l+k)^2} = \int\limits_0^1 dx\int\frac{d^4k}{(2\pi)^4}\frac{1}{R_{k-xl}\big(k^2+x(1-x)l^2\big)^2}\nonumber\\
&&\qquad\qquad\qquad\qquad\qquad\qquad\quad = \int\limits_0^1 dx \int\frac{d^4k}{(2\pi)^4}\frac{1}{R_{k}\big(k^2+x(1-x)l^2\big)^2} + O(\Lambda^{-2}),\qquad
\end{eqnarray}

\noindent
where we took into account that

\begin{equation}
R_{k-xl}\equiv R\Big((k_\mu-x l_\mu)^2/\Lambda^2\Big) = R(k^2/\Lambda^2) + O(\Lambda^{-2}) = R_k + O(\Lambda^{-2}).
\end{equation}

\noindent
Next, it is convenient to change the integration variable to $z= k^2/\Lambda^2$ and introduce the notation $\epsilon\equiv x(1-x)\,l^2/\Lambda^2$. Taking into account that $\epsilon$ is small, we obtain

\begin{eqnarray}
&&\hspace*{-7mm} \int \frac{d^4k}{(2\pi)^4} \frac{1}{R_{k}\big(k^2+x(1-x)l^2\big)^2} = \frac{1}{(4\pi)^2} \int\limits_0^{\infty}\frac{k^2dk^2}{R_k\big(k^2+x(1-x)l^2\big)^2} = \frac{1}{(4\pi)^2}\int\limits_0^{\infty}\frac{z dz}{R(z)(z+\epsilon)^2} \nonumber\\
&&\hspace*{-7mm} = \frac{1}{(4\pi)^2} \frac{\partial}{\partial \epsilon}\int\limits_0^{\infty} \frac{dz\, \epsilon}{R(z)(z+\epsilon)} = \frac{1}{(4\pi)^2}\frac {\partial}{\partial \epsilon}\bigg(\epsilon \ln(z+\epsilon)\frac{1}{R(z)}\biggr|_0^{+\infty}-\int\limits_0^{\infty}dz\, \epsilon \ln(z+\epsilon)\, \frac{d}{dz}\frac{1}{R(z)}\bigg)\nonumber\\
&&\hspace*{-7mm} = \frac{1}{(4\pi)^2} \frac{\partial}{\partial \epsilon} \bigg(-\epsilon\ln\epsilon-\int\limits_0^{\infty}dz\ln z\frac{d}{dz}\frac{\epsilon}{R(z)} + O(\epsilon^2\ln\epsilon)\bigg)
= \frac{1}{(4\pi)^2}\Big(-\ln\epsilon-1-A\Big) + O(\epsilon\ln\epsilon),\nonumber\\
\end{eqnarray}

\noindent where the constant $A$ is defined by Eq. (\ref{AB}). Integrating this expression with respect to $x$ and omitting terms vanishing in the limit $\Lambda\to\infty$, the considered integral can be written as

\begin{equation}
\int\frac{d^4k}{(2\pi)^4}\frac{1}{k^2R_k(l+k)^2}=\frac{1}{8\pi^2}\Big(\ln\frac{\Lambda}{l}+\frac{1}{2}-\frac{A}{2}\Big).
\end{equation}

\noindent The result for the integral $I_3$ is calculated by substituting this expression into Eq. (\ref{I3}),

\begin{equation}
I_3 = \frac{1}{128\pi^4}(1-A).
\end{equation}

According to Ref. \cite{Kazantsev:2018nbl},

\begin{eqnarray}
&& \lim\limits_{k\to 0} \int\frac{d^4l}{(2\pi)^4} \frac{1}{k^2 (k+l)^2 F_l}\Big(\frac{1}{F_{k+l}} - \frac{1}{F_l}\Big) = \lim\limits_{k\to 0}\bigg(\int\frac{d^4l}{(2\pi)^4} \frac{F_l-F_{k+l}}{(l^2-(k+l)^2)} \frac{1}{F_l^2 F_{k+l} (k+l)^2}\qquad\nonumber\\
&& - \int\frac{d^4l}{(2\pi)^4} \frac{F_l-F_{k+l}}{(l^2-(k+l)^2)} \frac{1}{F_l^2 F_{k+l} l^2}\bigg) = 0,
\end{eqnarray}

\noindent
because the last two integrals are well defined in the limit $k\to 0$. Therefore,

\begin{equation}
I_4 = I_5 = \lim\limits_{k\to 0}\, \frac{1}{8\pi^2} \Big( \int\frac{d^4l}{(2\pi)^4} \frac{1}{l^2F_l^2 (l+k)^2} - \frac{1}{8\pi^2} \ln\frac{\Lambda}{k} \Big).
\end{equation}

\noindent This expression is completely analogous to the integral $I_3$. Repeating the calculation described above, we obtain

\begin{equation}
I_4 = I_5 = \frac{1}{128\pi^4}(1-B),
\end{equation}

\noindent
where the constant $B$ is defined by Eq. (\ref{AB}).

Thus, the integrals (\ref{I1_Definition}) --- (\ref{I12_Definition}) are given by the following expressions:

\begin{eqnarray}
&&\hspace*{-7mm} I_1 =I_2 = \frac{1}{8\pi^2};\qquad\qquad\qquad\quad I_3 = \frac{1}{128\pi^4}(1-A); \qquad\qquad\quad I_4 = \frac{1}{128\pi^4}(1-B);\nonumber\\
&&\hspace*{-7mm} I_5 = \frac{1}{128\pi^4}(1-B);\qquad\qquad\quad I_6 = -\frac{1}{32\pi^4}\Big(\ln\frac{\Lambda}{\mu} + g_{12}\Big);\qquad\, I_7 = -\frac{1}{32\pi^4}\Big(\ln\frac{\Lambda}{\mu} + g_{11}\Big);\nonumber\\
&&\hspace*{-7mm} I_8 = \frac{A}{128\pi^4};\qquad\qquad\qquad\qquad\ I_9 = \frac{B}{128\pi^4};\qquad\qquad\qquad\qquad\ I_{10} = -\frac{1}{128\pi^4};\nonumber\\
&&\hspace*{-7mm} I_{11} = -\frac{3}{128\pi^4}\Big(\ln a_\varphi+1\Big);\qquad I_{12} = \frac{1}{128\pi^4}\Big(\ln a+1\Big).
\end{eqnarray}

\noindent
Substituting them into Eq. (\ref{Two_Loop_Gamma_Integrals}) the anomalous dimension can be written as

\begin{eqnarray}
&&\hspace*{-7mm} (\gamma_\phi)_i{}^j = -\frac{\alpha}{\pi}C(R)_i{}^j + \frac{1}{4\pi^2} \lambda^*_{imn}\lambda^{jmn} + \frac{\alpha^2}{2\pi^2} \left[C(R)^2\right]_i{}^j
+\frac{3\alpha^2}{2\pi^2}C_2C(R)_i{}^j\Big[\ln\frac{\Lambda}{\mu} + b_{11} - \ln a_\varphi\nonumber\\
&&\hspace*{-7mm} - 1 - \frac{A}{2}\Big] - \frac{\alpha^2}{2\pi^2} T(R) C(R)_i{}^j \Big[\ln\frac{\Lambda}{\mu} + b_{12} - \ln a - 1 - \frac{A}{2}\Big] -\frac{\alpha}{4\pi^3}\lambda^*_{lmn}\lambda^{jmn}C(R)_i{}^l\Big[\ln\frac{\Lambda}{\mu}+g_{11}\nonumber\\
&&\hspace*{-7mm} +\frac{1}{2}-\frac{B}{2}+\frac{A}{2} \Big]
- \frac{\alpha}{2\pi^3} \lambda^*_{imn}\lambda^{jml} C(R)_l{}^n \Big[ \ln\frac{\Lambda}{\mu} + g_{11} - \frac{1}{2}-\frac{B}{2}+\frac{A}{2} \Big]
+ \frac{1}{16\pi^4} \lambda^*_{iab}\lambda^{kab}\lambda^*_{kcd}\lambda^{jcd} \nonumber\\
&&\hspace*{-7mm} \times \Big[ \ln\frac{\Lambda}{\mu} + g_{12} \Big]
+\frac{1}{8\pi^4} \lambda^*_{iac}\lambda^{jab}\lambda^*_{bde}\lambda^{cde} \Big[ \ln\frac{\Lambda}{\mu} + g_{12} - \frac{1}{2}\Big] + O\Big(\alpha^3,\alpha^2\lambda^2,\alpha\lambda^4,\lambda^6\Big).
\end{eqnarray}

\noindent
Rewriting the result in terms of the bare couplings using Eqs. (\ref{Two_Loop_Alpha}) and (\ref{One_Loop_Lambda}) we obtain the anomalous dimension (\ref{Gamma_Bare}). Note that all $\ln\Lambda/\mu$ and all finite constants cancel each other in the resulting expression. Certainly, this can be considered as a correctness check.

\end{document}